\documentclass[article,aps,twocolumn,showpacs,amsmath,amssymb,superscriptaddress,nofootinbib,floatfix]{revtex4-2}

\usepackage{graphicx}
\usepackage{dcolumn}
\usepackage{bm}
\usepackage[colorlinks=true,linkcolor=blue,citecolor=blue,urlcolor=blue]{hyperref}

\usepackage{array,booktabs,tabularx}
\usepackage[caption=false]{subfig}
\usepackage[USenglish]{babel}
\usepackage{braket}
\usepackage{xcolor}
\usepackage{amsfonts}
\usepackage{amssymb}
\usepackage[normalem]{ulem}
\usepackage{amsmath}

\begin{document}

\title{Cascade of pressure induced competing CDWs in the kagome metal FeGe}

\author{A. Korshunov}
\thanks{These authors contributed equally to this work.}
\affiliation{Donostia International Physics Center (DIPC), Paseo Manuel de Lardizábal. 20018, San Sebastián, Spain}

\author{A. Kar}
\thanks{These authors contributed equally to this work.}
\affiliation{Donostia International Physics Center (DIPC), Paseo Manuel de Lardizábal. 20018, San Sebastián, Spain}

\author{C.-Y. Lim}
\thanks{These authors contributed equally to this work.}
\affiliation{Donostia International Physics Center (DIPC), Paseo Manuel de Lardizábal. 20018, San Sebastián, Spain}

\author{D. Subires}
\thanks{These authors contributed equally to this work.}
\affiliation{Donostia International Physics Center (DIPC), Paseo Manuel de Lardizábal. 20018, San Sebastián, Spain}
\affiliation{Departamento de Física Aplicada I, Universidad del País Vasco UPV/EHU, E-20018 San Sebastián, Spain}

\author{J. Deng}
\affiliation{Department of Applied Physics, Aalto University School of Science, FI-00076 Aalto, Finland}

\author{Y. Jiang}
\affiliation{Donostia International Physics Center (DIPC), Paseo Manuel de Lardizábal. 20018, San Sebastián, Spain}

\author{H. Hu}
\affiliation{Donostia International Physics Center (DIPC), Paseo Manuel de Lardizábal. 20018, San Sebastián, Spain}

\author{D. C\u{a}lug\u{a}ru}
\affiliation{Department of Physics, Princeton University, Princeton, NJ 08544, USA}

\author{C. Yi}
\affiliation{Max Planck Institute for Chemical Physics of Solids, 01187 Dresden, Germany}

\author{S. Roychowdhury}
\affiliation{Max Planck Institute for Chemical Physics of Solids, 01187 Dresden, Germany}
\affiliation{Department of Chemistry, Indian Institute of Science Education and Research Bhopal, Bhopal-462 066, India}

\author{C. Shekhar}
\affiliation{Max Planck Institute for Chemical Physics of Solids, 01187 Dresden, Germany}

\author{G. Garbarino}
\affiliation{European Synchrotron Radiation Facility (ESRF), BP 220, F-38043 Grenoble Cedex, France}

\author{P. T\"{o}rm\"{a}}
\affiliation{Department of Applied Physics, Aalto University School of Science, FI-00076 Aalto, Finland}

\author{C. A. Fuller}
\affiliation{Swiss-Norwegian BeamLines at European Synchrotron Radiation Facility}

\author{C. Felser}
\affiliation{Max Planck Institute for Chemical Physics of Solids, 01187 Dresden, Germany}

\author{B. Andrei Bernevig}
\affiliation{Donostia International Physics Center (DIPC), Paseo Manuel de Lardizábal. 20018, San Sebastián, Spain}
\affiliation{Department of Physics, Princeton University, Princeton, NJ 08544, USA}
\affiliation{IKERBASQUE, Basque Foundation for Science, 48013 Bilbao, Spain}

\author{S. Blanco-Canosa}
\email{sblanco@dipc.org}
\affiliation{Donostia International Physics Center (DIPC), Paseo Manuel de Lardizábal. 20018, San Sebastián, Spain}
\affiliation{IKERBASQUE, Basque Foundation for Science, 48013 Bilbao, Spain}

\begin{abstract}
Electronic ordering is prevalent in correlated systems, which commonly exhibit competing interactions. Here, we use x-ray diffraction to demonstrate a cascade of pressure induced order-disorder transformations, with propagation vectors \textbf{q}$_\mathrm{CDW}$= $\left(\frac{1}{2}\ 0\ \frac{1}{2}\right)$, \textbf{q}$^*$=$\left(\frac{1}{3}\ \frac{1}{3}\ \frac{1}{2}\right)$ and \textbf{q}$^\dag$=$\left(\frac{1}{3}\ \frac{1}{3}\ \frac{1}{3}\right)$, in the kagome metal FeGe. In the pressure interval between 4$<$\textit{p}$<$10 GPa, \textbf{q}$_\mathrm{CDW}$ and \textbf{q}$^*$ coexist and the spatial extent of the $\sqrt{3}\times\sqrt{3}$ order is nearly long-range at $\sim$15 GPa, $\sim$30 unit cells. Above $\sim$25 GPa, the periodic lattice distortion has a propagation wavevector of \textbf{q}$^\dag$=$\left(\frac{1}{3}\ \frac{1}{3}\ \frac{1}{3}\right)$ at room temperature. The cascade of phase transitions are captured by the Ising model of frustrated triangular lattices and modeled by Monte Carlo simulations based on the dimerization of trigonal Ge$_1$. The pressure dependence of the integrated intensities and correlation lengths of \textbf{q}$_\mathrm{CDW}$, \textbf{q}$^*$ and \textbf{q}$^\dag$ demonstrate a competition between the 2$\times$2 and $\sqrt{3}\times\sqrt{3}$ phases prove the tunability of order-disorder phase transitions under pressure and the rich landscape of metastable/fragile phases of FeGe.  


\end{abstract}

\maketitle

Correlated phases are usually characterized by a complex interplay of charge, spin, orbital and lattice degrees of freedom with comparable energy scales that allow for a fine tuning of their ground state by external stimuli \cite{Tsunetsugu_2009,Fulde,Dagotto2005}. Notable examples can be found in the gate-tuned superconductivity in SrTiO$_3$ \cite{Caviglia2008}, Mott insulators \cite{Imada1998} or twisted bilayer graphene (TBG) \cite{Cao2018}. Another example is the ground state of superconducting cuprates that exhibits an interplay of competing charge, spin and superconducting phases \cite{Frano_2020,Ghiringhelli_2012,Blanco_2013}, highly tunable under pressure \cite{Yamamoto2015,Souliou_2014}, strain \cite{Kim_2018,Vinograd2024,Bluschke2018} and magnetic fields \cite{Gerber_2015,Chang2016}.   

Very recently, a cascade of correlated phases have been reported in the geometrically frustrated kagome lattices \cite{Ortiz_2019,Ortiz_2020,Ronny_PRB_2012,Ronny_PRL_2013,Yin2022}, believed to be rooted on their rich electronic band structure, featuring van Hove singularities (vHs), Dirac crossings (DC) and dispersionless flat bands \cite{Kang_2020,Kang_2022,Yu_2022,Peng2021}. Such electronic details could unveil novel unconventional electrodynamics, like chiral CDWs, superconductivity or possible chiral flux phases \cite{Mielke2022,Guguchia2023,Guo2022}, which can be realized upon tuning the band fillings to the Fermi level. For instance, the non magnetic kagome metal AV$_3$Sb$_5$ (A=Cs, K, Rb) is characterized by a multiple-\textbf{q} CDW connecting the M and L points of the Brillouin zone (BZ) \cite{Miao_2021,Miao_2021_PRB}, intertwined with a low temperature superconducting state \cite{Felser_2022,Chen_2021}. Such 3D CDW is strongly influenced by the dimensionality \cite{Song2023} and stacking sequence \cite{Jin_2024}, magnetic field \cite{Jiang2021}, dynamic disorder \cite{Subires_2023}, uniaxial strain \cite{Guo2024,Qian2021} and pressure \cite{Tsirlin2023,Chen2021_pressure}. Intriguingly, the application of hydrostatic pressure results in the emergence of new electronic orders \cite{Yu2021,stier2024} that demonstrate the tunability of the AV$_3$Sb$_5$ ground state to external parameters. Similarly, the long range 3D CDW in the double layer ScV$_6$Sn$_6$ is characterized by a $\left(\frac{1}{3}\ \frac{1}{3}\ \frac{1}{3}\right)$ propagation vector \cite{Arachchige_2022} that competes with the high temperature short range charge fluctuations with wavevector $\left(\frac{1}{3}\ \frac{1}{3}\ \frac{1}{2}\right)$ \cite{Korshunov_2023,Cao_2023}, featuring a soft mode that becomes stable under moderate hydrostatic pressure \cite{Yi_2024}.

\begin{figure*}
    \centering
    \includegraphics[width=\linewidth]{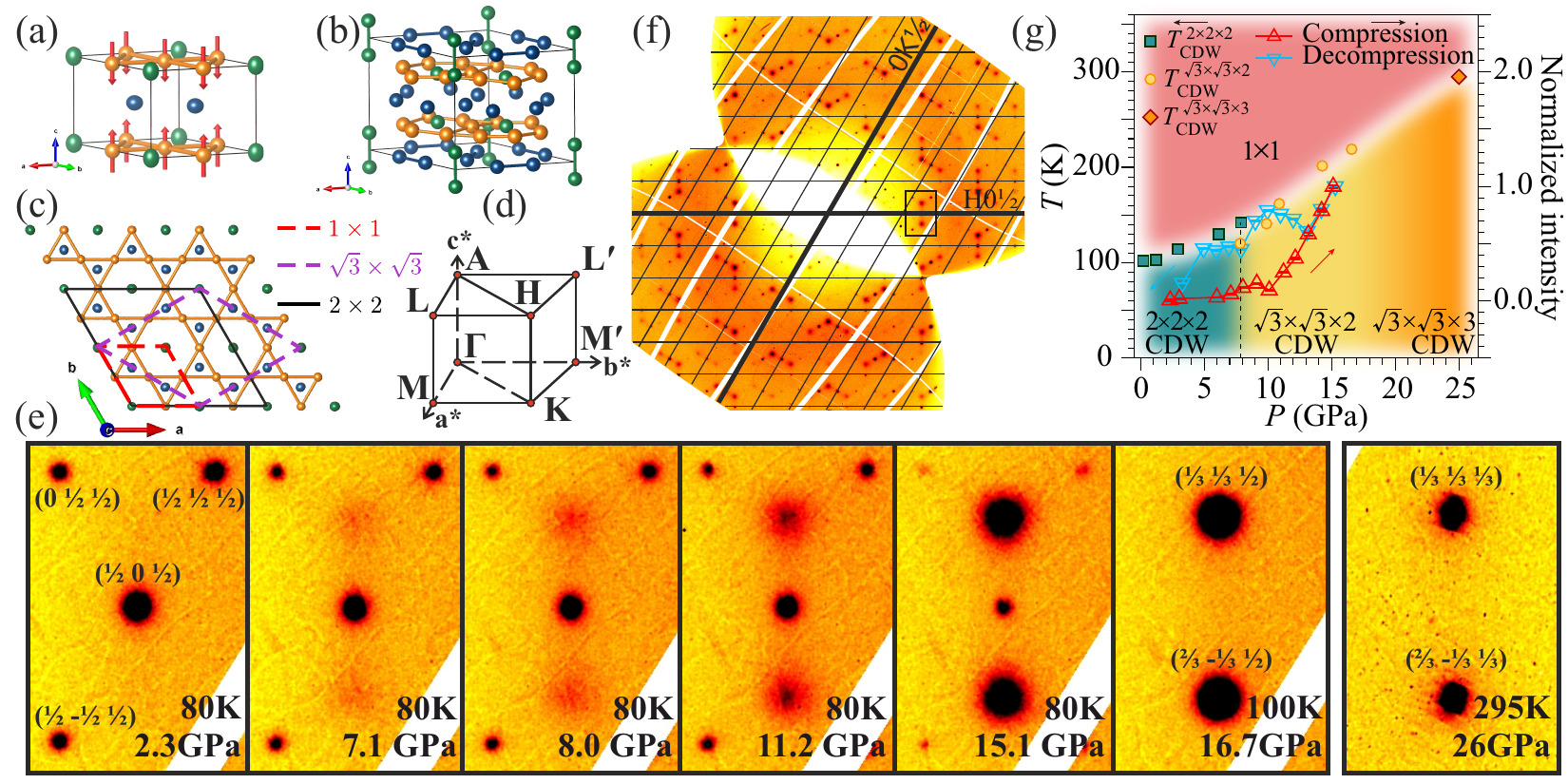}
    \caption{(a) Normal state structure (non-CDW) of FeGe. Arrows denote the spin polarization of the kagome planes, which will double the paramagnetic unit cell in the $z$-direction. Orange symbols are the Fe atoms, the green represents the Ge atoms in the kagome plane (trigonal Ge$_1$) and the blue symbols are the Ge atoms in the honeycomb layer (Ge$_2$). (b) Unit cell of the $\sqrt{3}\times\sqrt{3}$ CDW phase. (c)  Top view of the FeGe kagome net, displaying the 1$\times$1, 2$\times$2 and $\sqrt{3}\times\sqrt{3}$ superlattices. (d) High symmetry points in the non-magnetic Brillouin zone. (e) Detailed (h\ k\ $\frac{1}{2}$) maps parsing the pressure evolution of the different CDW orders of FeGe around the reciprocal lattice vector \textbf{G}$_\mathrm{200}$. (f) Reciprocal space map (RSM) sweeping the (h\ k\ $\frac{1}{2}$) plane at 80 K and 8 GPa, highlighting the DS that characterizes the crossover from the 2$\times$2 towards the $\sqrt{3}\times\sqrt{3}$ order. (g) Pressure-temperature (\textit{p}-T) phase diagram of FeGe (closed symbols). Superimposed (open symbols), compression (red)-decompression (blue) pressure hysteresis (see text).}  
    \label{Fig1}
\end{figure*}
 
Kagome FeGe crystallizes in the same space group as AV$_3$Sb$_5$ and ScV$_6$Sn$_6$ (\textit{P6/mmm}). However, FeGe orders antiferromagnetically below T$_\mathrm{N}$$\sim$400 K and develops a conical order at lower temperature \cite{Lebech_1987}. At T$_\mathrm{CDW}$$\sim$ 105 K, diffraction experiments unveil a dimerization driven short-range charge density wave  \cite{Teng_2022,Teng_2023,Yin_2022,Chen_2023,Teng_2024,Wang_2023, Zhao_2023, chen2023long} that connect the M, L, and A high symmetry points of the BZ. The microscopic origin of the CDW is still under debate, as angle resolved photoemission (ARPES) and density functional theory (DFT) show saddle points at the M point close to the Fermi level \cite{Teng_2023,Oh2024}, while scattering techniques report a giant spin-phonon coupling above T$_\mathrm{CDW}$ \cite{Miao_2023} and a moderate phonon softening below T$_\mathrm{CDW}$ \cite{Teng_2022}. Moreover, DFT calculations indicate that both the real and imaginary parts of the electronic susceptibility peak at the $\left(\frac{1}{3}\ \frac{1}{3}\ 0\right)$ wavevector (K point in the 2D BZ) \cite{Wu_2023,Yi_2023, wu2023electron}, in contrast to the propagation vectors observed experimentally. Finally, diffuse scattering (DS) locates the CDW within a canonical order-disorder transformation \cite{subires2024}, thus hinting at the tunability of the electronic order \cite{Chen_2023} to external perturbations. 

Given the metastability of the 2 $\times$ 2 phase of FeGe and its short-range correlation length, here we apply hydrostatic pressure to tune a cascade of transformations from the 2 $\times$ 2 towards a $\sqrt{3}\times\sqrt{3}\ \times$\ 2, \textbf{q}$^*$=$\left(\frac{1}{3}\ \frac{1}{3}\ \frac{1}{2}\right)$, and $\sqrt{3}\times\sqrt{3}\ \times$\ 3, \textbf{q}$^\dag$=$\left(\frac{1}{3}\ \frac{1}{3}\ \frac{1}{3}\right)$, periodic lattice distortion (PLD). We find that the onset temperature of the 2 $\times$ 2 order increases linearly with pressure up to nearly 150 K. Moreover, the intensity of \textbf{q}$_{CDW}$ decreases up to \textit{p}$^*$=15 GPa, above which, the nearly long-range $\sqrt{3}\times\sqrt{3}\ \times$\ 2 order develops. At $\sim$25 GPa, the $\sqrt{3}\times\sqrt{3}\ \times$\ 3 PLD becomes the ground state at room temperature and demonstrates a strong competition between the three CDW phases. DFT calculations confirm the stability of the $\sqrt{3} \times \sqrt{3}$ phase at high \textit{p} as a result of the dimerization of the trigonal Ge$_1$, that is well captured by the Ising model of frustrated triangular lattices and corroborates the extreme fragility of the electronic orders of FeGe. 


 
Figures \ref{Fig1}(a-b) sketch the unit cell of FeGe in its normal (NS), 2 $\times$ 2 and $\sqrt{3} \times \sqrt{3}$ state of FeGe, respectively. The NS unit cell consists on a 2D FeGe kagome layer and hexagonal Ge$_2$ layers, featuring a honeycomb lattice. At \textit{p}=0, the trigonal Ge$_1$ in the kagome plane dimerizes along the \textit{c}-direction, renormalizing the $p_z$-derived electron pocket at $\Gamma$ below T$_\mathrm{CDW}$ \cite{Teng_2024}, hence the 2$\times$2 PLD is a result of the frustrated order driven by the dimerization of Ge$_1$, as observed by x-ray diffraction \cite{Miao_2023,Shi_2023} and DFT calculations \cite{Wang_2023}. 
At 0 GPa, the CDW in FeGe is described by three independent $\mathrm{q}$-vectors, namely: $\left(\frac{1}{2}\ 0\ \frac{1}{2}\right)$ (L point), $\left(0\ 0\ \frac{1}{2}\right)$ (A point) and $\left(\frac{1}{2}\ 0\ 0\right)$ (M point) \cite{Teng_2022,Miao_2023}, see fig. \ref{Fig1}(d). Focusing on the \textbf{G}$_\mathrm{200}$ reciprocal lattice vector, fig. \ref{Fig1}(e), the intensity of the CDW peak with propagation vector $\left(\frac{1}{2}\ 0\ \frac{1}{2}\right)$ shows a gradual weakening up to 15 GPa. Similar pressure dependence of the intensity of the 2$\times$2 order is observed for $\left(\frac{1}{2}\ 0\ 0\right)$ (M point) and $\left(0\ 0\ \frac{1}{2}\right)$ (A point). 
In the pressure interval between 5 and 10 GPa, broad diffuse satellites emerge at $\left(\frac{1}{3}\ \frac{1}{3}\ \frac{1}{2}\right)$ and $\left(\frac{2}{3}\ -\frac{1}{3}\ \frac{1}{2}\right)$ (H point) and gain spectral weight up to \textit{p}$^*$$\sim$15 GPa. Above \textit{p}$^*$, the 2 $\times$ 2 order is reduced and falls below the detection limit.
Overall, our experimental results up to 25 GPa are summarized in the \textit{p}T-phase diagram presented in the fig. \ref{Fig1}(g). The onset of the CDW with propagation vector \textbf{q}$^*$=$\left(\frac{1}{3}\ \frac{1}{3}\ \frac{1}{2}\right)$ follows the same pressure dependence as the 2 $\times$ 2 order at a rate of $\Delta$T/\textit{p}$\sim$7 K/GPa, presumably due to the release of the dimerization driven lattice frustration \cite{subires2024}, in marked contrast with the AV$_3$Sb$_5$ and ScV$_6$Sn$_6$ kagome metals \cite{Chen2021_pressure,stier2024}. 
Further increase of pressure leads to the disappearance of the $\sqrt{3}\times\sqrt{3}\times2$ phase and a tripling of the unit cell along the \textit{c}-direction ($\sqrt{3} \times\sqrt{3} \times$ 3 PLD) at $\sim$25 GPa and room temperature, \textbf{q}$^\dag$=$\left(\frac{1}{3}\ \frac{1}{3}\ \frac{1}{3}\right)$.



\begin{figure}
    \centering
    \includegraphics[width=\linewidth]{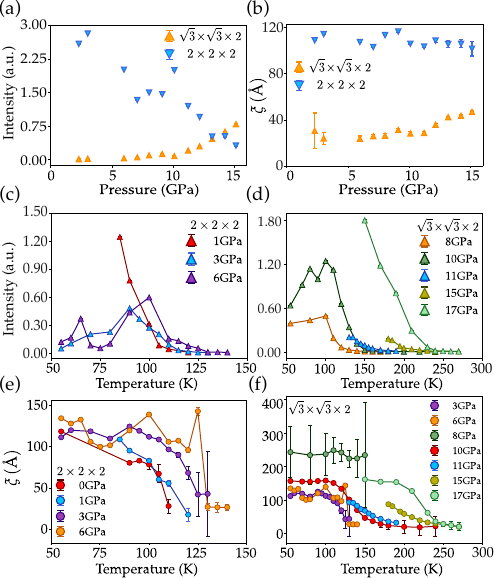}
    \caption{(a) Pressure dependence of the integrated intensities and correlation lengths (b) of the 2 $\times$ 2 and $\sqrt{3} \times \sqrt{3}$ orders at 80 K. (c) Temperature dependence of the intensity of the 2$\times$2 and (d) $\sqrt{3} \times \sqrt{3}\times$ 2 phases at different values of pressure. (e) Temperature dependence of the correlation length of the $2\times$2 and (f) $\sqrt{3}\times\sqrt{3}\times$ 2 orders at selected pressures. }
    \label{Fig2}
\end{figure}

To further delve into the evolution of the 2 $\times$ 2 and $\sqrt{3} \times \sqrt{3}$ orders as a function of \textit{p}, we parse the x-ray reciprocal space maps (RSM), see suppl. inf. fig. 3. From the pressure dependence of the 2$\times$2 and $\sqrt{3}\times\sqrt{3}\times2$ integrated intensities at T=80 K (where the integrated intensities reach their maximum value), we infer a mutual competition between both orders, fig. \ref{Fig2}(a). On the other hand, the correlation length of the 2$\times$2 order remains practically unaffected by pressure ($\sim$100 \r{A}), while the $\sqrt{3} \times\sqrt{3} \times$ 2 phase, yet less correlated, increases with \textit{p}, fig. \ref{Fig2}(b). Furthermore, the spatial distribution of the $\sqrt{3} \times \sqrt{3} \times$ 2 phase extends to $\sim$30 unit cells in the \textit{ab}-plane at low temperature (T$<$50 K) and 17 GPa, fig. \ref{Fig2} (f), hinting at a quasi-long-range order at high \textit{p}. 

\begin{figure*}
    \centering
    \includegraphics[width=0.85\linewidth]{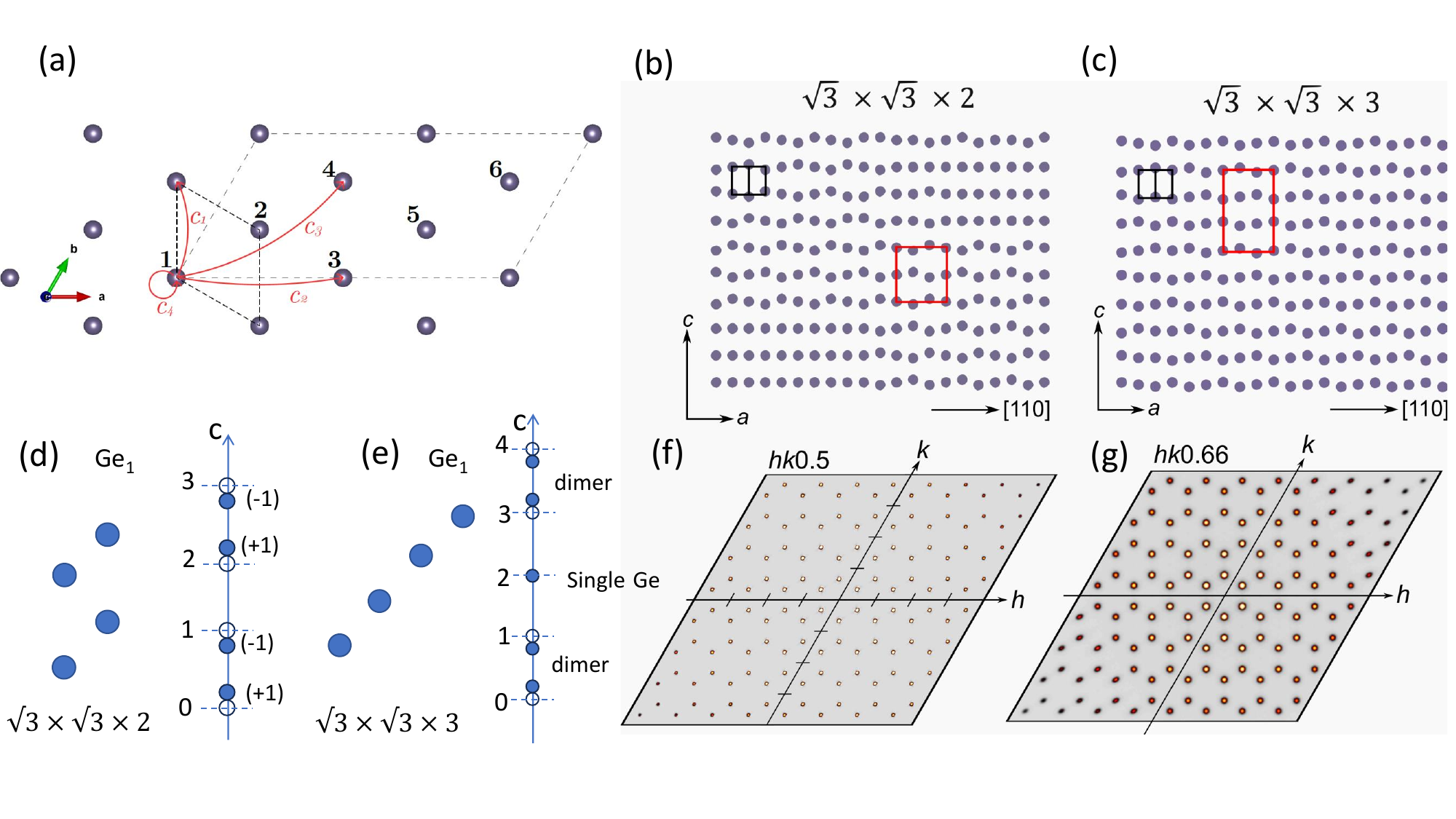}
    \caption{(a) Sketch of the model used for the MC simulations, where the interaction parameters of the Ising Hamiltonian are described as \textit{c}$_{i's}$.
    (b) Atomic arrangement of the trigonal Ge$_1$ atoms derived from the MC simulations using the \textit{ab initio} \textit{c}$_{i's}$ for the $\sqrt{3}\ \times\ \sqrt{3}\ \times$ 2 order. (c) Atomic configuration of the for the $\sqrt{3}\ \times\ \sqrt{3}\ \times$ 3 order. The red squares highlight the two-fold and three-fold superlattice along the \textit{c}-direction. (d) and (e) describe the dimerization arrangement along the \textit{c}-direction for the for the $\sqrt{3}\ \times\ \sqrt{3}\ \times$ 2 and for the $\sqrt{3}\ \times\ \sqrt{3}\ \times$ 3 orders, respectively. +1 (-1) stands for the upward (downward) displacement of the Ge$_1$. (f) and (g) Fourier transform of 
    (b) and (c), respectively, showing the $\frac{1}{3}\ \frac{1}{3}\ \frac{1}{2}$ and $\frac{1}{3}\ \frac{1}{3}\ \frac{1}{3}$ superlattice peaks.}
    \label{Fig3}
\end{figure*}

Notably, once the pressure is cranked up, the temperature dependence of the integrated intensity at constant \textit{p} of the 2 $\times$ 2 and $\sqrt{3} \times\sqrt{3} \times$ 2 superlattice reflections start to peak right below the CDW onset and further weakens down to 10 K, fig. \ref{Fig2}(c-d). Similar behavior has been observed in FeGe$_{0.9}$ crystals \cite{subires2024}, presumably associated to quenched disorder or the internal strain induced by the Ge deficiency, and in correlated systems with competing/coexisting orders \cite{Ricci_2021,Blanco_2013}. On the other hand, the temperature dependence of the correlation lengths for both orders, fig. \ref{Fig2}(e-f) follows a gradual decrease up to T$_\mathrm{CDW}$ due to either the formation of dislocations or to a shrinking of the ordered domains into smaller regions. Intriguingly, upon pressure cycling (compression-decompression below T$_\mathrm{CDW}$), the 2 $\times$ 2 order can only be restored after warming up above T$_\mathrm{CDW}$ and subsequent cooling, indicating a extreme fragility of 2$\times$2 phase to pressure, thus the $\sqrt{3} \times\ \sqrt{3}$ phase is the ground state of FeGe at high \textit{p} and a metastable phase at low \textit{p}, fig. \ref{Fig1}(g).

\begin{figure}
    \centering
    \includegraphics[width=\linewidth]{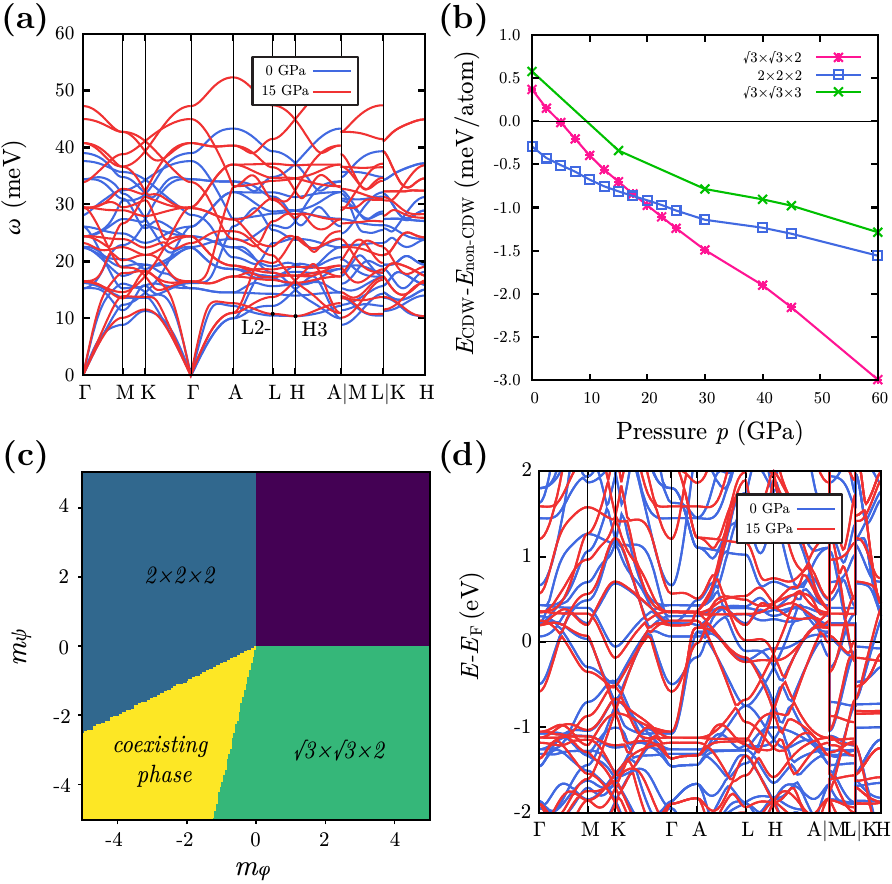}
    \caption{(a) Phonon spectra at 0 and 15 GPa. Note that the phonon spectra are calculated in the antiferromagnetic phase and plotted in the paramagnetic BZ. (b) The difference of total energy between the CDW structure and non-CDW structure in the antiferromagnetic phase versus \textit{p} for $\sqrt{3} \times \sqrt{3}$ and 2$\times$2 orders. (c) Phase diagram derived from the Landau-Ginzburg free energy. (d) Band structure of antiferromagnetic FeGe at 0 and 15 GPa.}
    \label{Fig4}
\end{figure}

Given that the $\sqrt{3} \times \sqrt{3}\times$ 2 PLD evolves from the 2 $\times$ 2 order, featuring and order-disorder transformation of dimerized Ge$_1$ \cite{subires2024}, we model the $\sqrt{3} \times \sqrt{3} \times$ 2 order following the Ising scenario of antiferromagnetic triangular lattices, fig. \ref{Fig3}(a) (suppl. inf. Appendix D), where the strength and direction of interacting spins is controlled by the \textit{c}$_i$ coefficients in the eq. \ref{MC}. 

\begin{equation}
\begin{split}
    H = \sum_{<i,j>:NN}c_1\sigma_i\sigma_j + \sum_{<i,j>:NNN}c_2\sigma_i\sigma_j +  \\ \sum_{<i,j>:4NN}c_3\sigma_i\sigma_j + \sum_{<i,j>:z-NN}c_4\sigma_i\sigma_j + \sum_{i}h\sigma_i + E_0,
    \end{split}
\label{MC}
\end{equation}

The \textit{c}$_i$ parameters were obtained \textit{ab initio} and take the values \textit{c}$_1$= 32.495 meV, \textit{c}$_2$=-17.68 meV, \textit{c}$_3$= 5.166 meV and \textit{c}$_4$= -14.232 meV. The MC derived atomic configuration of the $\sqrt{3}\ \times\ \sqrt{3}\ \times$ 2 order is plotted in fig. \ref{Fig3}(b) with patches of \textit{zig-zag} dimerization along the \textit{c}-direction, fig. \ref{Fig3}(d). The Fourier transform of such Ge$_1$ pattern is displayed in fig. \ref{Fig3}(f) and nicely reproduces the experimental x-ray diffraction pattern. 

On the other hand, the $\sqrt{3}\ \times\ \sqrt{3}\ \times$ 3 order requires a modified Ising hamiltonian, since the dimers need to be considered as 2 distinct Ge$_1$ atoms, suppl. inf. Appendix D. Moreover, thee model requires a \textit{c}$_5$ coefficient in the eq. \ref{MC} that accounts for the second nearest neighbor interaction along the \textit{c}-direction.\begin{equation}
\sum_{<i,j>:z-NNN}c_5\sigma_i\sigma_j,
\end{equation} 
The MC simulation reproduces the atomic pattern displayed in fig. \ref{Fig3}(c), where the Ge$_1$ dimerizes diagonally along the \textit{c}-direction, fig. \ref{Fig3}(e). Remarkably, the Fourier transform of such Ge$_1$ configuration, presented in fig. \ref{Fig3}(g), shows Bragg spots at the ($\frac{1}{3}\ \frac{1}{3}\ \frac{1}{3}$) wavevector, again in good agreement with the experimental data at $\sim$25 GPa, hence, the MC simulations capture in full detail the cascade of order-disorder transitions in FeGe under hydrostatic pressure.
 
Microscopically, understanding the pressure dependence of the CDW in FeGe represents a challenge. As shown in fig. \ref{Fig4}(a), the phonon spectra remain stable within the harmonic approximation and most of the phonon modes harden with \textit{p} (suppl. inf. fig. 8). However, the phonon at the H point, characterized by the irreducible representation (irrep) $H3$, remains nearly constant in energy. This $H3$ phonon corresponds to the dimerization of triangular Ge$_1$ atoms of the $\sqrt{3} \times \sqrt{3}$ order, similarly as the lowest phonon at the L point (irrep $L2-$) at low pressure, fig. \ref{Fig4}(a), (suppl. inf. fig. 8). 
We have carried out DFT calculations to investigate the total energy of different CDW structures at various pressures. As shown in  fig. \ref{Fig4}(b), the $2 \times 2$ CDW structure has lower energy, consistent with existing experimental results \cite{Teng_2022, Chen_2023}. However, as pressure increases, the total energy of the $\sqrt{3} \times \sqrt{3} \times$ 2 superstructure decreases more rapidly and becomes the ground state above 18 GPa, in good agreement with our experimental observations. On the other hand, the $\sqrt{3} \times \sqrt{3} \times$ 3 order presents higher energy at the DFT level, see fig. \ref{Fig4}(b), in contrast to the experimental observations, hinting at the possible crucial role of the lattice anharmonicity at high \textit{p}.  
The 2$\times$2 to $\sqrt{3} \times \sqrt{3} \times$ 2 phase transition can be theoretically described by the Landau-Ginzburg theory, which incorporates the coupling between the order parameters associated with the two distinct CDW phases, as described in suppl. inf. Appendix E.
The phase diagram obtained from Landau-Ginzburg analysis is presented in fig. \ref{Fig4}(c), where the two tuning parameters \(m_\phi\) and \(m_\psi\) represent the masses of the respective order parameters: $\phi$ for $2\times 2$ CDW and $\psi$ for $\sqrt{3} \times \sqrt{3} \times$ 2 CDW. This phase diagram illustrates that a phase transition occurs from a \(2 \times 2\) order to a phase where both CDW orders coexist, and eventually to a $\sqrt{3} \times \sqrt{3} \times$ 2 CDW phase at high \textit{p}. 

The rotation of the in-plane propagation vector and the tripling of the \textit{c}-axis with \textit{p}, although confirmed by DFT and MC simulations, is a remarkable result, which deserves further analysis. The calculated band structures of antiferromagnetic FeGe at $p=0$ and 15 GPa are shown in (suppl. inf. fig. 6) fig. \ref{Fig4}(d). Roughly, as pressure increases, the van Hove singularities at the M and L points shift away from the Fermi energy E$_\mathbf{F}$, while the electron pockets at the K point expand with \textit{p}. This expansion does not enhance the Fermi surface nesting at different K points with respect to the low \textit{p} phase, as shown by the non-interacting charge susceptibility plot, $\chi_\mathrm{\textbf{q}}$ (suppl. inf. fig. 7). Nevertheless, the peak in $\chi_\mathrm{\textbf{q}}$ at K suggests that charge correlations might be central in the emergence of the $\sqrt{3} \times \sqrt{3}$ CDW order. 
We would like to finish with a brief discussion about the weakening of the CDW intensities of the $\sqrt{3} \times \sqrt{3}\ \times$ 2 order below T$_\mathrm{CDW}$. One possibility might be related to the local deformations induced by the large dimerization of the Ge$_1$ atoms that could act as strong pinning centers of the mobile carriers leading to the formation of standing waves, i.e., Friedel oscillations (FOs) \cite{DallaTorre2016,White2002,Rouziere2000}. The modulations observed might simply be due to FOs around local sources of disorder that screen the local distortions and stabilize the charge fluctuations near T$_\mathrm{CDW}$ at 2\textit{k}$_F$. Below T$_\mathrm{CDW}$, the quasi-long-range order might suppress the disorder and gives rise to a drop of the $\sqrt{3} \times \sqrt{3}$ intensity \cite{Yue2020}, as we have observed in fig. \ref{Fig3}. 


In summary, we have carried out a high-pressure x-ray diffraction and DFT study in the antiferromagnetic kagome metal FeGe that reveals quasi-long-range $\sqrt{3} \times \sqrt{3}$ order with propagation vectors \textbf{q}$^*$=$\left(\frac{1}{3}\ \frac{1}{3}\ \frac{1}{2}\right)$ and \textbf{q}$^\dag$=$\left(\frac{1}{3}\ \frac{1}{3}\ \frac{1}{3}\right)$, that compete with the low-pressure 2$\times$2 ordered phase. The $\sqrt{3} \times \sqrt{3}$ orders are described by a large dimerization of the trigonal Ge$_1$ atoms in the kagome plane and modeled by Monte Carlo simulations based on the Ising hamiltonian of frustrated triangular lattices and demonstrate the strong tunability of the fragile ground state of FeGe under pressure. 

\textit{Note added}: During the completion of this manuscript, we became aware of related results reporting a possible high pressure $\sqrt{3} \times \sqrt{3} \times$ 6 phase in annealed FeGe crystals by Xikai Wen et al. \cite{Wen2024}.  

\section{Acknowledgments}
We acknowledge Ming Yi and Stephen Wilson for fruitful discussions. D.S., A.Kar and S.B-C. acknowledge financial support from the MINECO of Spain through the project PID2021-122609NB-C21 and by MCIN and by the European Union Next Generation EU/PRTR-C17.I1, as well as by IKUR Strategy under the collaboration agreement between Ikerbasque Foundation and DIPC on behalf of the Department of Education of the Basque Government. A.K. thanks the Basque government for financial support through the project PIBA-2023-1-0051.   
J.D. and P.T. were supported by Jane and Aatos Erkko Foundation, Keele Foundation, and Magnus Ehrnrooth Foundation as part of the SuperC collaboration. J. D. acknowledges the computational resources provided by the Aalto Science-IT project. B.A.B., C.F. and P.T. belong to the SuperC collaboration.
Y.J. and H.H. were supported by the European Research Council (ERC) under the European Union’s Horizon 2020 research and innovation program (Grant Agreement No. 101020833) as well as by IKUR Strategy. 
D.C\u{a}l. acknowledges the hospitality of the Donostia International Physics Center, at which this work was carried out. 
D.C\u{a}l. and C.-Y.L. were supported by the European Research Council (ERC) under the European Union’s Horizon 2020 research and innovation program (grant agreement no. 101020833) and by the Simons Investigator Grant No. 404513. B.A.B was supported by the Gordon and Betty Moore Foundation through Grant No.GBMF8685 towards the Princeton theory program, the Gordon and Betty Moore Foundation’s EPiQS Initiative (Grant No. GBMF11070), Office of Naval Research (ONR Grant No. N00014-20-1-2303), Global Collaborative Network Grant at Princeton University, BSF Israel US foundation No. 2018226, NSF-MERSEC (Grant No. MERSEC DMR 2011750), Simons Collaboration on New Frontiers in Superconductivity and the Schmidt Foundation at the Princeton University.

\bibliography{FeGe}

\end{document}


\title{Supplementary information for `Cascade of pressure induced competing CDWs in the kagome metal FeGe'}

\maketitle
\appendix
\tableofcontents

\section{Experimental section}
\subsection{Single crystal growth and technical details}

Single crystals of FeGe were grown by the chemical vapor transport method, as described elsewhere \cite{subires2024,Wenzel_2024}. High-pressure single crystal diffraction experiments were carried out at the ID15b beamline of the European Synchrotron Radiation Facility (ESRF). The sample was loaded in a diamond anvil cell (DAC) using He as the pressure medium. The pressure was calibrated by using ruby luminescence. We used an incident energy E$_\mathrm{i}$=30 keV (0.41 Å) and EIGER2 X 9M CdTe detector to collect the scattered photons. The observed Bragg reflections were indexed using CrysAlisPro (Rigaku Oxford Diffraction) software, which was used to obtain reciprocal space maps. The components (h\ k\ l) of the scattering vector are expressed in reciprocal lattice units (r.l.u.), (h\ k\ l)= $h \mathbf{a}^*+k \mathbf{b}^*+l \mathbf{c}^*$, where $\mathbf{a}^*$, $\mathbf{b}^*$, and $\mathbf{c}^*$ are the reciprocal lattice vectors of the paramagnetic-normal state unit cell. 

\subsection{\label{app:cryst}Crystal Structure and Symmetry}

The crystal structure of FeGe in the A-AFM phase exhibits the symmetry of a type-IV Shubnikov space group (SG), specifically the magnetic space group (MSG) 192.252 $P_c6/mcc$ in the BNS setting, or MSG 191.13.1475 $P_{2c}6/mm'm'$ in the OG setting. According to the OG setting convention, where symmetry operations are expressed in the original nonmagnetic unit cell, FeGe possesses an anti-unitary translation $\mathcal{T}\cdot\{E|001\}$ and a unitary halving subgroup $P 6/mcc$. The anti-unitary translation $\mathcal{T}\cdot\{E|001\}$ connects two Kagome layers and reverses the spin direction, resulting in spin degeneracy within the A-AFM band structure [in the A-AFM Brillouin zone (BZ)], as depicted in \cref{fig:el-pressure}.

\begin{figure}
    \centering
    \includegraphics[width=1.0\linewidth]{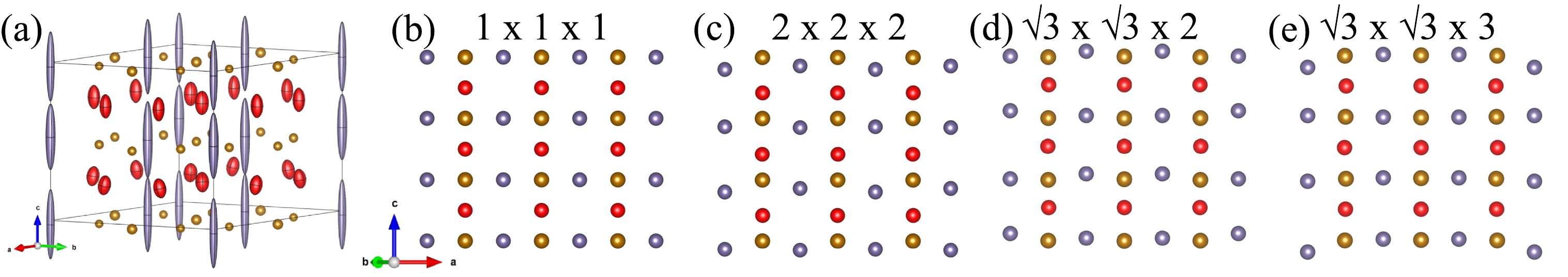}
    \caption{(a) Refined average crystal structures of FeGe in the $\sqrt{3}\ \times\ \sqrt{3}\ \times$ 2 PLD. (b) 1 $\times$ 1 (c) 2 $\times$ 2 (d) $\sqrt{3}\ \times\ \sqrt{3}\ \times$ 2 and (e) $\sqrt{3}\ \times\ \sqrt{3}\ \times$ 3 superlattices. Orange circles stand for Fe, grey circles represent the Ge atoms in the kagome plane (trigonal Ge$_1$) and red circles are the Ge atoms in the honeycomb layer (Ge$_2$).}
    \label{fig:cryst-struc}
\end{figure}

The crystal structures for the non-CDW phase, as well as the $\sqrt{3}\times \sqrt{3}\times$ 3, $\sqrt{3}\times \sqrt{3}\times$ 3 and $2\times 2\times 2$ CDW phases, are illustrated in \cref{fig:cryst-struc}{(b) and (e)} respectively, with the supercell in the CDW phases referenced relative to the paramagnetic (PM) phase. In the CDW phases, the Kagome layers of Fe and the triangular layers of Ge are positioned near the $z=0$ and $z=\flatfrac{\vb{a}_3}{2}$ planes. The most significant atomic displacements induced by the CDW orders are the movements of the triangular Ge atoms along the $z$-axis, as highlighted in \cref{fig:cryst-struc} (a). These displacements correspond to an effective phonon orbital situated at $(0, 0, \flatfrac{1}{4})$.

\begin{table}[h!]
\centering
\caption{Crystal data and structure refinement for the modulated \(\sqrt{3} \times \sqrt{3} \times 2\) superlattice measured at 100 K and 16.7 GPa: \(P6mm\) symmetry (Space group: 183). Unit cell dimensions: \(a = b = 8.3810(6)\, \text{\AA}\), \(c = 7.784(6)\, \text{\AA}\), \(\alpha = \beta = 90^\circ\), \(\gamma = 120^\circ\). Volume: \(473.5\, \text{\AA}^3\). \(R_f = 5.89\%\), GOF = \(1.69\%\).}
\label{tab:FeGe_332_structure}
\begin{tabular}{cccccccc}
\toprule
Atom & \(x\) & \(y\) & \(z\) & Occ. & \(U\) & Site & Symm. \\
\midrule
Ge1   & 0.0000 & 0.0000 & -0.0284 & 1 & 0.088 & 1a  & 6mm \\
Ge2   & 0.3333 & 0.6667 &  0.0128 & 1 & 0.109 & 2b  & 3m. \\
Ge3   & 0.3371 & 0.0000 &  0.7488 & 1 & 0.009 & 6d  & ..m \\
Ge4   & 0.3301 & 0.0000 &  0.2483 & 1 & 0.009 & 6d  & ..m \\
Ge5   & 0.0000 & 0.0000 &  0.5228 & 1 & 0.087 & 1a  & 6mm \\
Ge6   & 0.3333 & 0.6667 &  0.4843 & 1 & 0.103 & 2b  & 3m. \\
Fe1   & 0.5000 & 0.5000 & -0.0231 & 1 & 0.006 & 3c  & 2mm \\
Fe2   & 0.5000 & 0.5000 &  0.4722 & 1 & 0.005 & 3c  & 2mm \\
Fe3   & 0.3333 & 0.1667 & -0.0282 & 1 & 0.004 & 6e  & .m. \\
Fe4  & 0.3330 & 0.1667 &  0.4741 & 1 & 0.005 & 12f & 1 \\
\bottomrule
\end{tabular}
\end{table}

\begin{table}[h!]
\centering
\caption{Crystal data and structure refinement for the modulated \(\sqrt{3} \times \sqrt{3} \times 3\) superlattice measured at RT and 26.7 GPa: \(P6mm\) symmetry (Space group: 183). Unit cell dimensions: \(a = b = 14.345(1)\, \text{\AA}\), \(c = 11.629(7)\, \text{\AA}\), \(\alpha = \beta = 90^\circ\), \(\gamma = 120^\circ\). Volume: \(2072.4\, \text{\AA}^3\). \(R_f = 16.2\%\), GOF = \(8.3\%\).}
\label{tab:FeGe_333_structure}
\begin{tabular}{cccccccc}
\toprule
Atom & \(x\) & \(y\) & \(z\) & Occ. & \(U\) & Site & Symm. \\
\midrule
Ge1   &  0.11260 &  0.22530 &  0.14600 & 1 & 0.020 & 6e  & .m. \\
Ge2   & -0.11180 & -0.22350 & -0.19100 & 1 & 0.017 & 6e  & .m. \\
Ge3   &  0.10780 &  0.21560 &  0.47700 & 1 & 0.013 & 6e  & .m. \\
Ge4   &  0.10770 &  0.55390 &  0.14500 & 1 & 0.019 & 6e  & .m. \\
Ge5   & -0.10930 & -0.55460 & -0.19000 & 1 & 0.016 & 6e  & .m. \\
Ge6   &  0.11740 &  0.55870 &  0.47500 & 1 & 0.013 & 6e  & .m. \\
Ge7   &  0.44080 &  0.22040 &  0.14500 & 1 & 0.019 & 6e  & .m. \\
Ge8   & -0.44280 & -0.22140 & -0.19000 & 1 & 0.017 & 6e  & .m. \\
Ge9   &  0.45080 &  0.22540 &  0.47600 & 1 & 0.012 & 6e  & .m. \\
Ge10  &  0.00000 &  0.00000 &  0.07500 & 1 & 0.002 & 1a  & 6mm \\
Ge11  &  0.00000 &  0.00000 &  0.34600 & 1 & 0.035 & 1a  & 6mm \\
Ge12  &  0.00000 &  0.00000 & -0.29000 & 1 & 0.020 & 1a  & 6mm \\
Ge13  &  0.00000 &  0.33210 &  0.02800 & 1 & 0.014 & 6d  & ..m \\
Ge14  &  0.00000 &  0.33210 &  0.38500 & 1 & 0.006 & 6d  & ..m \\
Ge15  &  0.00000 & -0.33300 & -0.29500 & 1 & 0.012 & 6d  & ..m \\
Ge16  &  0.33333 &  0.66667 &  0.07000 & 1 & 0.002 & 2b  & 3m. \\
Ge17  &  0.33333 &  0.66667 &  0.33000 & 1 & 0.010 & 2b  & 3m. \\
Ge18  & -0.33333 & -0.66667 & -0.29600 & 1 & 0.015 & 2b  & 3m. \\
Fe1   &  0.00000 &  0.16250 & -0.02200 & 1 & 0.013 & 6d  & ..m \\
Fe2   &  0.00000 &  0.17050 &  0.31200 & 1 & 0.014 & 6d  & ..m \\
Fe3   &  0.00000 & -0.17060 & -0.35000 & 1 & 0.004 & 6d  & ..m \\
Fe4   &  0.00000 &  0.50000 & -0.01300 & 1 & 0.003 & 3c  & 2mm \\
Fe5   &  0.00000 &  0.50000 &  0.32200 & 1 & 0.014 & 3c  & 2mm \\
Fe6   &  0.00000 & -0.50000 & -0.35200 & 1 & 0.016 & 3c  & 2mm \\
Fe7   &  0.33420 &  0.16710 & -0.01900 & 1 & 0.014 & 6e  & .m. \\
Fe8   &  0.33333 &  0.16667 &  0.31500 & 1 & 0.010 & 6e  & .m. \\
Fe9   & -0.33040 & -0.16520 & -0.35800 & 1 & 0.008 & 6e  & .m. \\
Fe10  &  0.33260 &  0.50190 & -0.01860 & 1 & 0.008 & 12f & 1 \\
Fe11  &  0.33590 &  0.50030 &  0.31410 & 1 & 0.006 & 12f & 1 \\
Fe12  & -0.33290 & -0.50010 & -0.35000 & 1 & 0.011 & 12f & 1 \\
\bottomrule
\end{tabular}
\end{table}

\subsection{Pressure dependence of the lattice parameters.}

\begin{figure}
    \centering
    \includegraphics[width=0.85\linewidth]{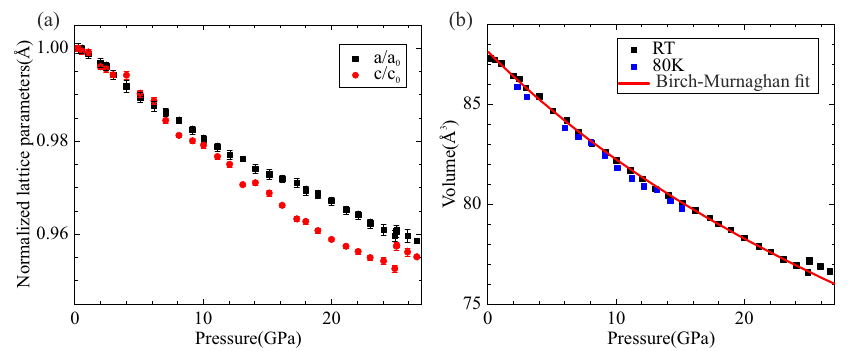}
    \caption{Pressure dependence of the lattice parameters and volume of FeGe and its fitting to the third order Birch-Murnaghan equation of state. Inset: Normalized pressure dependence of the lattice parameters, with the normalization value taken at low pressure.}
    \label{Fig2_SI}
\end{figure}

The pressure dependence of the lattice parameters of FeGe at 80 K is shown in Table \ref{tab:FeGe_Lattice} and plotted in fig. \ref{Fig2_SI}.

\begin{table}[h!]
\centering
\caption{Pressure dependence of lattice parameters and volume of FeGe in parent 1$\times$1$\times$1 P6/mmm cell measured at RT.}
\label{tab:FeGe_Lattice}
\begin{tabular}{cccc}
\toprule
$P$ (GPa) & $a$ (\AA) & $c$ (\AA) & $V$, (\AA$^3$) \\
\midrule
0.2  & 4.989(5) & 4.052(3) & 87.42 \\
0.5  & 4.988(5) & 4.049(3) & 87.30 \\
1.1  & 4.983(5) & 4.048(3) & 87.09 \\
2.0  & 4.973(5) & 4.036(3) & 86.49 \\
2.4  & 4.970(5) & 4.033(3) & 86.31 \\
3.0  & 4.960(5) & 4.029(3) & 85.90 \\
4.0  & 4.948(6) & 4.028(4) & 85.49 \\
5.1  & 4.936(6) & 4.012(4) & 84.81 \\
6.2  & 4.927(5) & 4.006(4) & 84.34 \\
7.1  & 4.920(4) & 3.989(3) & 83.70 \\
8.1  & 4.912(4) & 3.976(2) & 83.14 \\
9.2  & 4.901(4) & 3.971(2) & 82.70 \\
10.0 & 4.891(5) & 3.967(3) & 82.31 \\
11.1 & 4.883(5) & 3.957(3) & 81.81 \\
12.1 & 4.874(5) & 3.950(3) & 81.38 \\
13.1 & 4.870(2) & 3.933(1) & 80.81 \\
14.1 & 4.859(5) & 3.934(3) & 80.55 \\
15.1 & 4.853(5) & 3.925(3) & 80.16 \\
16.1 & 4.849(4) & 3.915(2) & 79.72 \\
17.3 & 4.845(5) & 3.903(2) & 79.28 \\
18.0 & 4.836(5) & 3.901(3) & 79.01 \\
18.9 & 4.832(4) & 3.893(2) & 78.67 \\
20.0 & 4.825(4) & 3.885(2) & 78.27 \\
21.1 & 4.815(4) & 3.879(2) & 77.87 \\
22.1 & 4.810(5) & 3.875(2) & 77.59 \\
23.0 & 4.801(5) & 3.869(3) & 77.21 \\
24.1 & 4.794(6) & 3.866(4) & 76.91 \\
24.9 & 4.787(6) & 3.860(4) & 76.57 \\
25.1 & 4.794(3) & 3.880(2) & 77.20 \\
25.0 & 4.792(7) & 3.879(4) & 77.13 \\
25.9 & 4.787(7) & 3.874(5) & 76.84 \\
26.6 & 4.782(3) & 3.870(2) & 76.62 \\
\bottomrule
\end{tabular}
\end{table}

Despite the 2D character of the kagome plane, we find a nearly isotropic contraction of both the \textit{a} and \textit{c}-axis, $\sim$2.5-3\%. Fitting the pressure dependence of the volume to the third order Birch-Murnaghan equation of state (EoS), 

\begin{equation}
\begin{aligned}
    P\left(V\right) = & \frac{3B_0}{2}\left[\left(\frac{V_0}{V}\right)^{7/3}-\left(\frac{V_0}{V}\right)^{5/3}\right]\\
    & \left\{1+\frac{3}{4}\left(B_0'-4\right)\left[\left(\frac{V_0}{V}\right)^{2/3}-1\right]\right\},
\end{aligned}
\end{equation}

where $\mathrm{P}$ is the pressure, $\mathrm{V}_0$ is the volume at zero pressure, $\mathrm{B}_0$ is the Bulk modulus, and $\mathrm{B}_0$´ is the derivative of the bulk modulus with respect to pressure, returns a bulk modulus of $\mathrm{B}_0 = 135\pm3\,\mathrm{GPa}$, larger than AV$_3$Sb$_5$ \cite{Yu_2022_pressure,Zhang_2021} and ScV$_6$Sn$_6$ \cite{Kong_2024}, V$_0$ = 87.7$\pm$0.1 and B$_{0'}$= 4.7$\pm$0.3.

\subsection{Analysis of the x-ray diffraction under pressure.}

\begin{figure}
    \centering
    \includegraphics[width=0.9\linewidth]{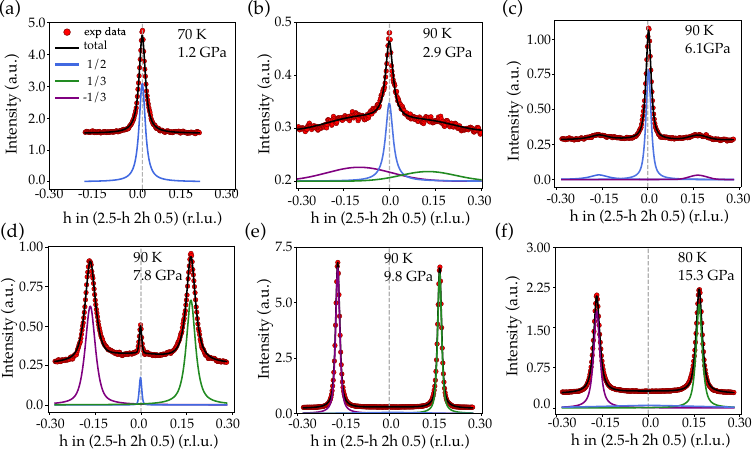}
    \caption{H-cuts of the RSM at selected pressures and temperatures, labeled inset. The charge peak with propagation vector ($\frac{1}{2}\ 0 \frac{1}{2}$) in (a) is gradually suppressed under pressure and replaced by satellites with wave vector ($\frac{1}{3}\ \frac{1}{3}\ \frac{1}{2}$) at 17 GPa in (f).}
    \label{Fig3_SI}
\end{figure}

To determine the transition temperature, we collected diffuse scattering data at several constant pressures for a sequence of temperatures close to the phase transition. 
The pressure dependence of both the 2$\times$2 and $\sqrt{3}\times\sqrt{3}$ orders was fitted to a pseudo-Voigt profile, a convolution of Gaussian and Lorentzian functions, fig. \ref{Fig3_SI}. Figures \ref{Fig3_SI} summarize the transition from the 2$\times$2 to the $\sqrt{3}\times\sqrt{3}$ phases with pressure.

\begin{figure}
    \centering
    \includegraphics[width=0.9\linewidth]{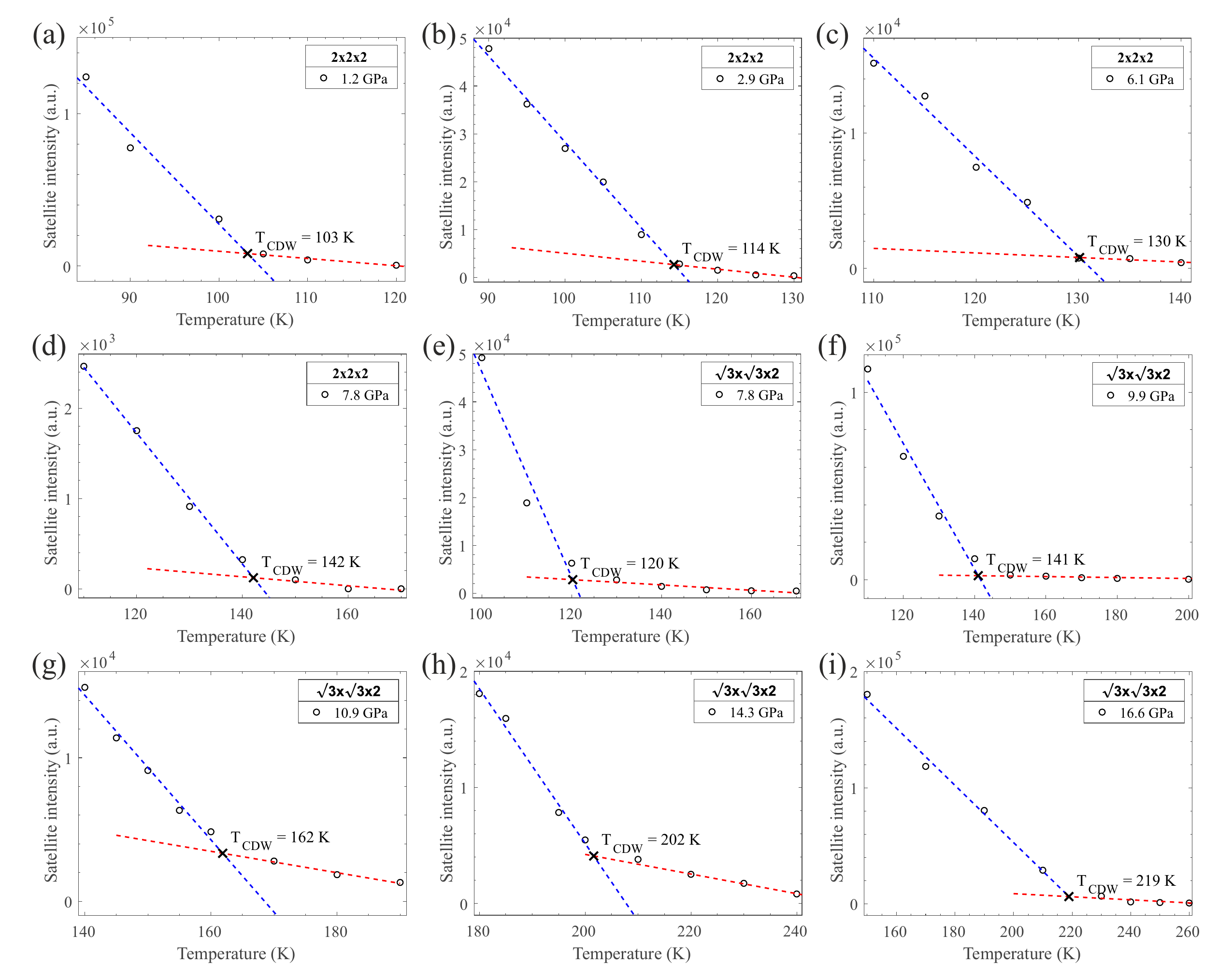}
    \caption{Determination of the onset of the T$_\mathrm{CDW}$.}
    \label{Fig_Trans_SI}
\end{figure}

The transition temperature for each pressure was determined in the same manner as the intersection of two linear dependencies: one for the low-temperature phase and the other for the high-temperature phase in the region of the phase transition, where a strong increase in diffuse intensity was observed, fig. \ref{Fig_Trans_SI}. During the temperature cycle, a small pressure drift occurred, which was continuously monitored using ruby fluorescence. To accurately determine the temperature-pressure point corresponding to the transition temperature, this drift was approximated by a linear dependence on the \textit{p}T graph.

\begin{figure}
    \centering
    \includegraphics[width=0.9\linewidth]{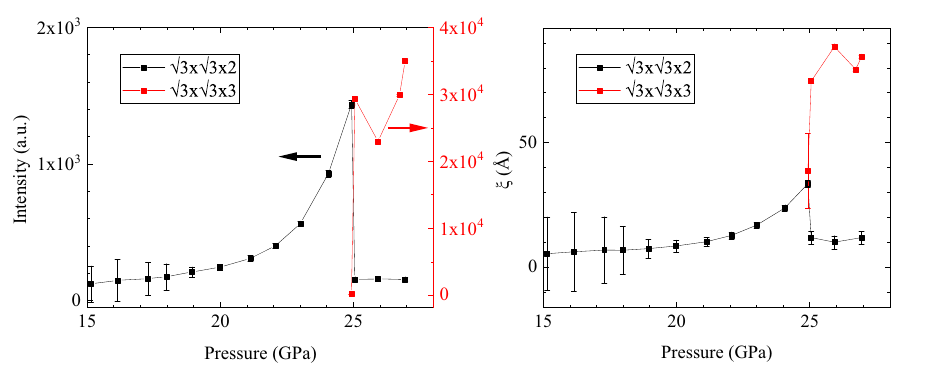}
    \caption{Pressure dependence of the intensity and linewidth of the $\sqrt{3}\ \times\ \sqrt{3}\ \times$ 2 and $\sqrt{3}\ \times\ \sqrt{3}\ \times$ 3 phases.}
    \label{Fig5_SI}
\end{figure}

On the other hand, at RT weak precursors of diffuse scattering (DS) corresponding to the $\sqrt{3}\ \times\ \sqrt{3}\ \times$ 2 phase were observed, which then abruptly transitioned to a new phase with a wave vector ($\frac{1}{3}\ \frac{1}{3}\ \frac{1}{3}$) at around 25 GPa. This transition was accompanied by a slight increase in lattice parameters: 0.52\% along the \textit{c}-direction and 0.14\% in the kagome plane, supporting the dimerization nature of the CDW phase. The intensity of $\sqrt{3}\ \times\ \sqrt{3}\ \times$ 2 phase immediately dropped, replaced by the scattering from $\sqrt{3}\ \times\ \sqrt{3}\ \times$ 3 phase at room temperature, fig. \ref{Fig5_SI}. Moreover, comparison of linewidth indicates that the $\sqrt{3}\ \times\ \sqrt{3}\ \times$ 3 phase occupies a larger lattice volume and represents the main ground state at high pressure.

The sharp crossover is indicative of a first order transition and the correlation length of the $\sqrt{3}\ \times\ \sqrt{3}\ \times$ 3 order at 30 GPa is 5 times the $\sqrt{3}\ \times\ \sqrt{3}\ \times$ 2 order at 15 GPa.  

\section{\label{app:dft}DFT Calculation Details}

First-principles calculations were performed based on density functional theory (DFT), using the projector augmented wave (PAW) method \cite{Blochl1994ProjectorA, Kresse1999FromA}, as implemented in the Vienna \textit{ab initio} Simulation Package (VASP) \cite{Kresse1996EfficiencyA, Kresse1996EfficientA}. The generalized gradient approximation (GGA) was used, incorporating the Perdew-Burke-Ernzerhof (PBE) exchange-correlation functional \cite{Perdew1996GeneralizedA}. The lattice parameters, $a$ and $c$, were relaxed in non-charge density wave (non-CDW) states by applying the desired pressure using the \texttt{PSTRESS} keyword. All calculations assumed the A-type antiferromagnetic (A-AFM) configuration. The internal atomic positions within the CDW superstructure were relaxed until the forces on each atom were reduced to less than $0.0001\;\text{eV}/\mathring{\text{A}}$. A kinetic energy cutoff of 500 eV was set for the plane-wave basis set. The Brillouin zone (BZ) was sampled using the Monkhorst-Pack method \cite{Monkhorst1976SpecialA} with a $\Gamma$-centered $k$-point grid of $12 \times 12 \times 9$ for the $2\times 2\times 2$ and $\sqrt{3}\times \sqrt{3}\times 2$ CDW superstructures. 

Phonon spectra were calculated using the frozen phonon method in conjunction with the \texttt{PHONOPY} package \cite{Togo2023ImplementationA, Togo2023First-principlesA}. Irreducible representations (irreps) were determined using \texttt{IRVSP} \cite{Gao2021IrvspA} and \texttt{WannierTools} \cite{Wu2018WannierToolsA}. The band representation (BR) analysis was conducted on the website \webUnconMat. Furthermore, a Wannier-based tight-binding Hamiltonian was constructed using \texttt{Wannier90} \cite{Mostofi2008Wannier90A, Mostofi2014An-updatedA, Pizzi2020Wannier90A}.

\section{\label{app:cdw}CDW Phase Transition Under Pressure}

In this section, we further explore the electronic and phononic properties of FeGe under pressure and discuss the potential emergence of CDW phases.


\subsection{\label{app:el-pressure}Electronic Structure Under Pressure}

\begin{figure}
    \centering
    \includegraphics[width=0.8\linewidth]{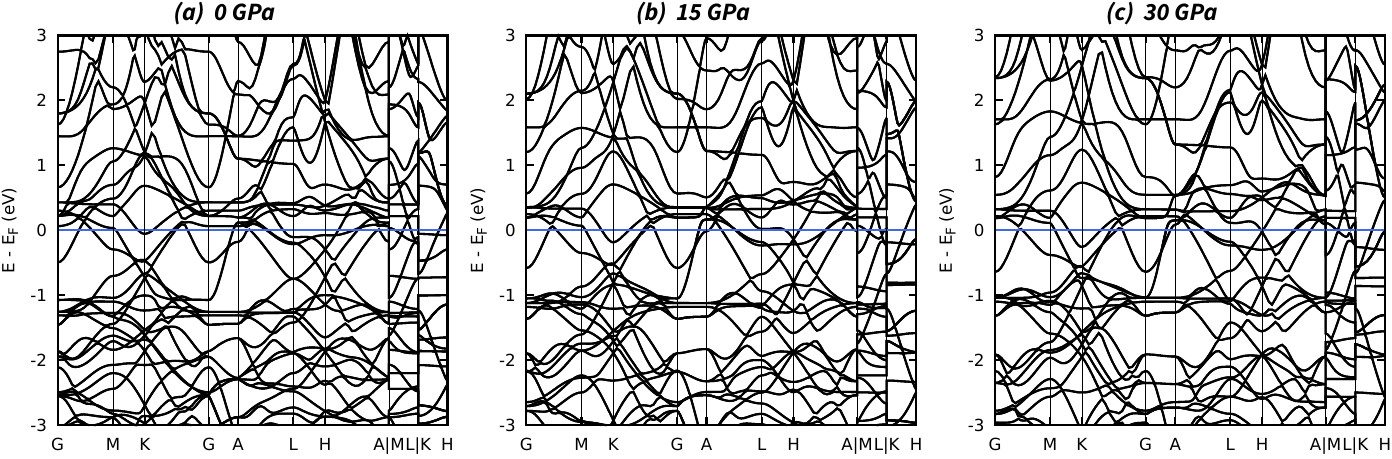}
    \caption{The band structures of A-AFM FeGe in non-CDW structure at (a) $P=0$, (b) 15, (c) 30 GPa, respectively.}
    \label{fig:el-pressure}
\end{figure}

\begin{figure}
    \centering
    \includegraphics[width=\linewidth]{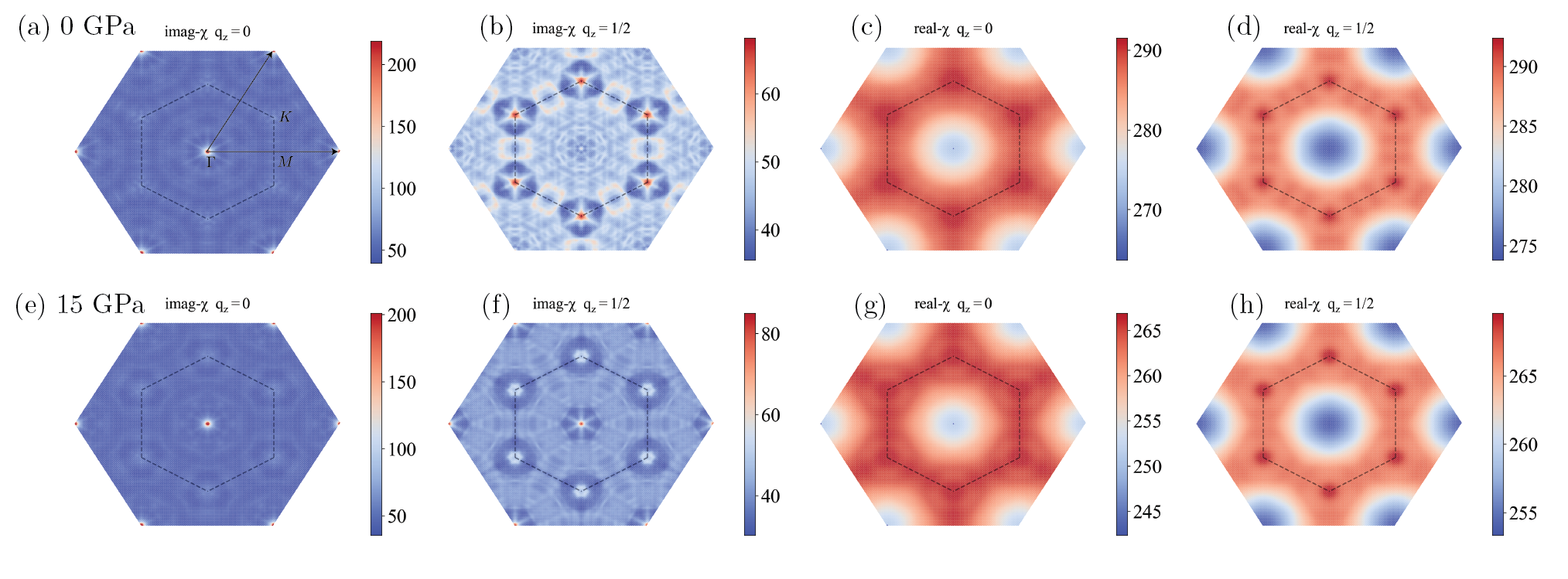}
    \caption{Fermi surface nesting function and charge susceptibility of FeGe in the A-AFM phase. (a)(b) are the nesting function on $k_z=0,\pi$ planes, respectively, and (c)(d) are the charge susceptibility at 0 GPa. (e)-(h) are the same but for 15 GPa. }
    \label{fig:nesting-suscept}
\end{figure}

First, we examined how the band structure of FeGe evolves under pressure. The calculated band structures of A-AFM FeGe in the non-CDW phase at $P = 0$, 15, and 30 GPa are shown in \cref{fig:el-pressure}{(a), (b), and (c)}, respectively. As pressure is applied, the van Hove singularities at the $M$ and $L$ points shift away from the Fermi energy $E_F$, and the electron pockets at the $K$ point expand with increasing pressure. This expansion of the Fermi surfaces at $K$ enhances the Fermi surface nesting effect due to the nesting of Fermi surfaces at different $K$ points, as indicated by the non-interacting charge susceptibility $\chi(\vb{q})$ computed from the Wannier-based tight-binding Hamiltonian of A-AFM FeGe, shown in \cref{fig:nesting-suscept}. This enhancement suggests that the emergent $\sqrt{3}\times \sqrt{3}\times 2$ CDW order could be driven by charge instability.

\subsection{\label{app:ph-pressure}Phononic Structure Under Pressure}

\begin{figure}
    \centering
    \includegraphics[width=0.8\linewidth]{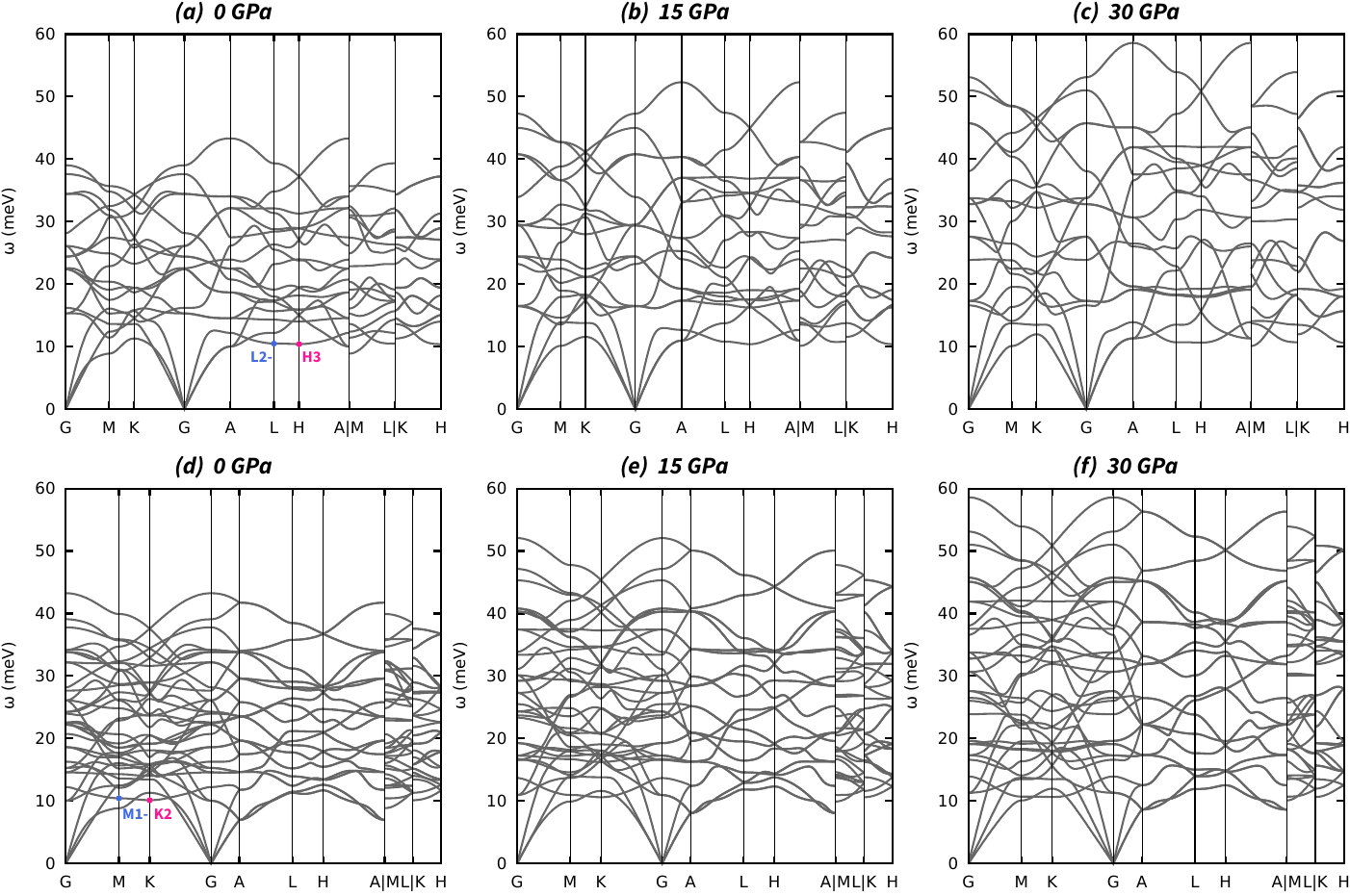}
    \caption{The phonon spectra of FeGe in non-CDW structure at (a)(d) $P=0$, (b)(e) 15, (c)(f) 30 GPa, respectively. The phonon spectra in (a), (b), and (c) are plotted in the nonmagnetic BZ of FeGe, while those in (d), (e), and (f) are plotted in AFM BZ.}
    \label{fig:ph-pressure}
\end{figure}

\cref{fig:ph-pressure} presents the phonon spectra of FeGe in the non-CDW state, calculated using DFT at several selected pressures. The phonon spectra of FeGe remain stable under pressure when analyzed within the harmonic approximation.
Energy calculations for various charge density wave (CDW) phases indicate that a CDW phase consistently has lower energy compared to the non-CDW phase as we shown in \cref{fig:en-diff} and discussed below. 

However, as most of the phonon modes harden with increasing pressure, a specific phonon at the $H$ point, characterized by the irreducible representation (irrep) $H3$, maintains nearly constant energy. This $H3$ phonon induces the dimerization of triangular Ge atoms located at the $2e$ Wyckoff position of SG 191. Additionally, the lowest phonon at the $L$ point, with irrep $L2-$, also displays dimerization characteristics. 

Starting from the phonon spectrum in the non-CDW AFM phase, these two irreps are utilized to generate the potential CDW structures corresponding to the $2\times 2\times 2$ and $\sqrt{3}\times \sqrt{3}\times 2$ phases. Since the most symmetric space group resulting from the condensation of these two phonons is SG 191 \cite{Jiang2023KagomeA}, we adhere to this space group when analyzing CDW structures. The atomic movements associated with the $H3$ and $L2-$ phonons are depicted in \cref{fig:cryst-struc}{(d) and (e)}.

To investigate the experimental CDW transitions, we generate the CDW structure from the lowest phonon modes at $L=(-1/2, 1/2, 1/2)$ and $H=(1/3, 1/3, 1/2)$, which correspond to the $2\times 2\times 2$ and $\sqrt{3}\times \sqrt{3}\times 2$ orders, respectively. As reported in previous literature \cite{Teng2022DiscoveryA, Jiang2023KagomeA, Wang2023EnhancedA, Wen2024UnconventionalA, Subires2024FrustratedA, Chen2024DiscoveryA, Chen2024InstabilityA}, the $2\times 2\times 2$ CDW transition is primarily driven by the significant dimerization of triangular Ge. This large dimerization is not fully captured by the lowest phonon modes at $L$ or $H$. Therefore, we increase the magnitude of the Ge dimerization in the generated CDW structure, ensuring that the symmetry remains intact, as the triangular Ge atoms are not at high-symmetry Wyckoff positions.

The dimerizations of triangular Ge atoms in the $\sqrt{3}\times \sqrt{3}\times 2$ and $2\times 2\times 2$ CDW phases are illustrated in \cref{fig:cryst-struc}{(d) and (e)}, respectively.



\subsection{\label{app:dimer}Triangular Ge dimerization}

\begin{figure}
    \centering
    \includegraphics[width=0.8\linewidth]{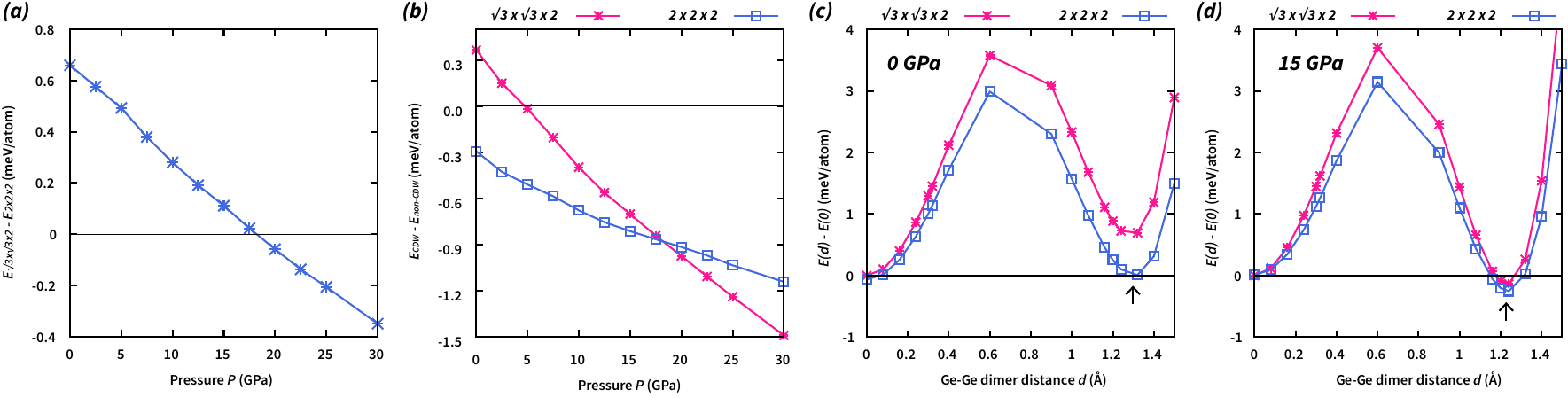}
    \caption{Energy curves.
    (a) The difference of total energy between the $\sqrt{3}\times \sqrt{3}\times 2$ and $2\times 2\times 2$ CDW structures with respect to pressure.
    (b) The difference of total energy between the CDW structure and non-CDW structure versus pressure.
    The difference of total energy between the CDW and non-CDW structure, $E = E_{\text{tot}}(d) - E_{\text{tot}}(d = 0)$, as functions of the distance $d$ between triangular Ge-dimerization and $c/2$, \ie $d=d_0-d_{\text{dimer}}$, at (c) $P=0$ GPa and (d) $P=15$ GPa. The black arrow labels a local energy minimum around $d= 1.3 \;\mathring{\text{A}}$ at 0 GPa, and $d= 1.2 \;\mathring{\text{A}}$ at 15 GPa}
    \label{fig:en-diff}
\end{figure}

To investigate the effect of triangular Ge dimerization on energy, we performed total energy calculations using DFT at different dimerization distances $d$ [as defined in \cref{fig:cryst-struc}(b)] under ambient pressure (AP) and high pressure (15 GPa, HP). As indicated by the local energy minima (shown by black arrows) in the energy curves in \cref{fig:en-diff}{(c) and (d)}, it can be observed that triangular Ge dimerization in the $2\times 2\times 2$ superstructure saves more energy than in the $\sqrt{3}\times \sqrt{3}\times 2$ superstructure at AP. However, as pressure increases, this energy difference diminishes.

To further investigate the CDW transition from an \textit{ab initio} perspective, we calculated the total energy of different CDW structures at various pressures. The total energies (relative to the non-CDW total energies) of fully relaxed structures with significant triangular Ge dimerization in the $\sqrt{3}\times \sqrt{3}\times 2$ and $2\times 2\times 2$ reconstructions at pressures up to 30 GPa are presented in \cref{fig:en-diff}{(b)}. At AP, the $2\times 2\times 2$ CDW structure has lower energy, consistent with existing experimental results \cite{Chen2024DiscoveryA, Subires2024FrustratedA}. However, as pressure increases, the energy of the $\sqrt{3}\times \sqrt{3}\times 2$ CDW structure decreases more rapidly, becoming the lower energy state around 18 GPa, as shown in \cref{fig:en-diff}{(a)}. This result is qualitatively consistent with experimental observations, which show a CDW phase transition at pressures up to 10 GPa, as discussed in the main text.

\begin{figure}
    \centering
    \includegraphics[width=0.8\linewidth]{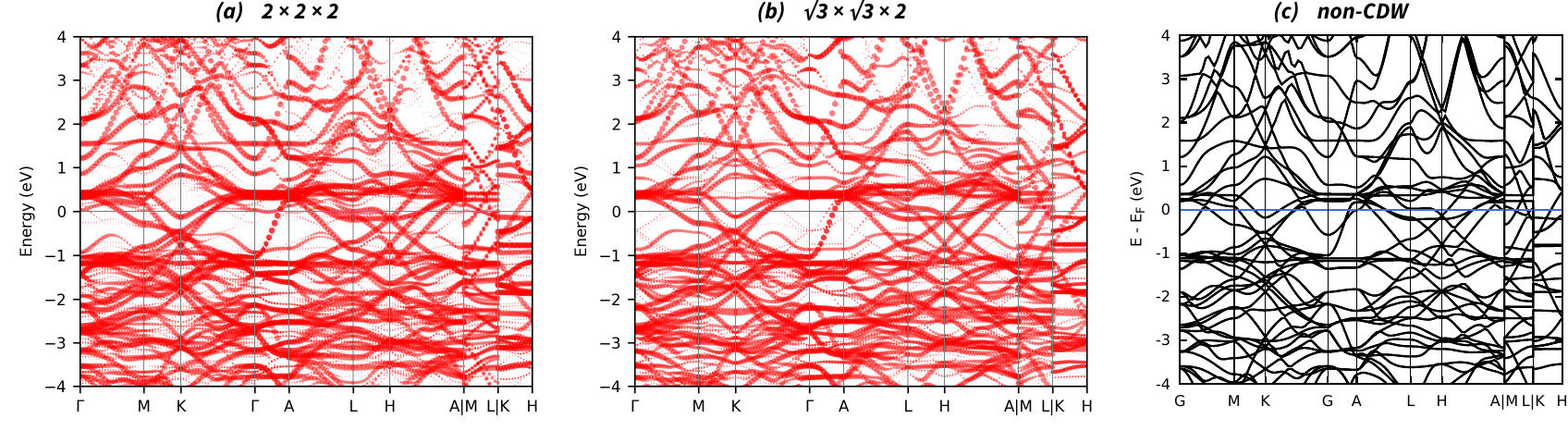}
    \caption{The unfolded band structures of FeGe in (a) $2\times 2 \times 2$ CDW, (b) $\sqrt{3}\times \sqrt{3}\times 2$ CDW, and (c) non-CDW structures at $P=15$ GPa.}
    \label{fig:unfold-el-hp}
\end{figure}

The unfolded band structures of different structures under high pressure are shown in \cref{fig:unfold-el-hp}, from which it can be seen that the CDW renormalization effect on the electronic structure is weak.

\subsection{Ising model for Ge dimerization}
Motivated by \cite{Subires2024FrustratedA}, to explain the rotation of CDW wave vector from $(1/2,0,1/2)$ to $(1/3,1/3,1/2)$, we first build an Ising model to describe the dimerization of triangular Ge and to achieve a certain microscopic realization via the Monte Carlo simulations. We use an Ising variable $\sigma_i = \pm 1$ to denote the dimerized ($\sigma_i = -1$) and undimerized ($\sigma_i = +1$) triangular Ge pair in the (non-CDW AFM phase) unit cell $\vb{R}_i$. Considering in-plane nearest neighbor (NN) coupling $c_1$, next NN (NNN) coupling $c_2$, 3rd NN (TNN) coupling $c_3$, $z$-direction NN (NNz) coupling $c_4$, and an effective magnetic field $h$, we build a model with the form
\begin{aleq}\label{eq:ising}
    H \eq \sum_{\ev{i j}_{\mathrm{NN}}} c_1 \sigma_i \sigma_j + \sum_{\ev{i j}_{\mathrm{NNN}}} c_2 \sigma_i \sigma_j + \sum_{\ev{i j}_{\mathrm{TNN}}} c_3 \sigma_i \sigma_j + \sum_{\ev{i j}_{\mathrm{NNz}}} c_4 \sigma_i \sigma_j + \sum_i h \sigma_i + E_0.
\end{aleq}
We then fit the parameters from DFT at both AP and HP. We choose a supercell of the non-CDW AFM unit cell, with reconstruction matrix
\begin{aleq}
    \pqty{\vb{a}', \vb{b}', \vb{c}'}^{T} \eq \mathcal{U} \pqty{\vb{a}, \vb{b}, \vb{c}}^{T}, \quad
    \mathcal{U} = \mqty( 4 & 2 & 0 \\ 1 & 2 & 0 \\ 0 & 0 & 2 ),
\end{aleq}
which contains $3 \times 2 \times 2$ Ge pairs. As shown in \cref{tab:ising_conf}, we consider 7 inequivalent dimer configurations and compute their averaged magnetic moments and total energies in DFT. The parameters in the Ising model \cref{eq:ising} are fitted using the DFT data, with their values summarized in \cref{tab:ising_para}.

\begin{table}[b!]
    \centering
    \caption{The computed total energies from DFT for different dimer configurations. For each dimer configuration, $+ (-)$ denotes the dimerized (undimerized) triangular Ge pair. The first $6$ $\pm$ denotes the 8 Ge pairs marked in \cref{fig:ising} on the first layer in the supercell, and the second $6$ $\pm$ denotes the second layer. The second (fifth) column of $\bar{\mu}_{\mathrm{Fe}}$ is the averaged magnitude of magnetic moment on Fe atoms at AP (HP). The third (sixth) column is the computed DFT total energy at AP (HP), while the forth (last) column is the fitted energy using \cref{eq:ising} at AP (HP). Note that the configurations with more dimerized Ge have larger magnetic moments and much higher total energies, and are not used in the fitting.}
    \label{tab:ising_conf}
    \begin{tabular}{c||c|c|c||c|c|c}
        \hline\hline
        & \multicolumn{3}{|c||}{AP}  & \multicolumn{3}{|c}{HP} \\
        \hline
        Dimer configuration & $\bar{\mu}_{\text{Fe}}/\mu_{\text{B}}$ & Total energy (meV) & Fitted energy (meV) & $\bar{\mu}_{\text{Fe}}/\mu_{\text{B}}$ & Total energy (meV) & Fitted energy (meV) \\
        \hline
        $++++++,++++++$ & $1.49$ & $626.5$ & $579.4$ & $1.25$ & $791.5$ & $711.9$ \\
        $--++++,--++++$ & $1.57$ & $747.9$ & $747.9$ & $1.33$ & $1109.6$ & $1061.8$ \\
        $-+-+++,-+-+++$ & $1.57$ & $500.6$ & $500.6$ & $1.33$ & $0$ & $-23.9$ \\
        $-+++++,-+++++$ & $1.53$ & $388.4$ & $388.4$ & $1.29$ & $435.9$ & $626.9$ \\
        $-++-++,-++-++$ & $1.57$ & $0$ & $0$ & $1.33$ & $973.6$ & $925.8$ \\
        $-+-+++,++++++$ & $1.53$ & $461.2$ & $555.4$ & $1.28$ & $457.9$ & $457.9$ \\
        \hline\hline
    \end{tabular}
\end{table}

\begin{table}[b!]
    \centering
    \caption{The fitted value of parameters in the Ising model \cref{eq:ising} based on the \textit{ab-initio} data in \cref{tab:ising_conf}. The parameters are in unit of meV.}
    \label{tab:ising_para}
    \begin{tabular}{c|c|c|c|c|c|c}
        \hline\hline
         & $c_1$ & $c_2$ & $c_3$ & $c_4$ & $h$ & $E_0$ \\
        \hline
        AP & $34.414$ & $18.950$ & $-6.618$ & $-33.217$ & $-247.75$ & $2809.6$ \\
        \hline
        HP & $32.495$ & $-17.680$ & $5.166$ & $-14.232$ & $-133.988$ & $1771.290$ \\
        \hline\hline
    \end{tabular}
\end{table}

\begin{figure}
    \centering
    \includegraphics[width=0.5\linewidth]{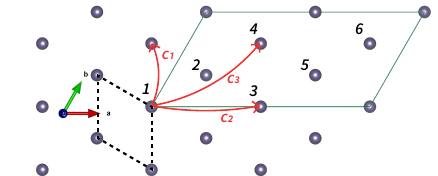}
    \caption{The illustration of the coupling parameters used in the effective Ising model in \cref{eq:ising}. $c_{1,2,3}$ are in-plane NN, NNN, and TNN coupling parameters. The numbers $1-6$ label the six Ge pairs in a in-plane supercell of the non-CDW unit cell. The black dashed lines mark the non-CDW unit cell.}
    \label{fig:ising}
\end{figure}

\subsection{Additional CDW}
\begin{figure}
    \centering
    \includegraphics[width=0.9\linewidth]{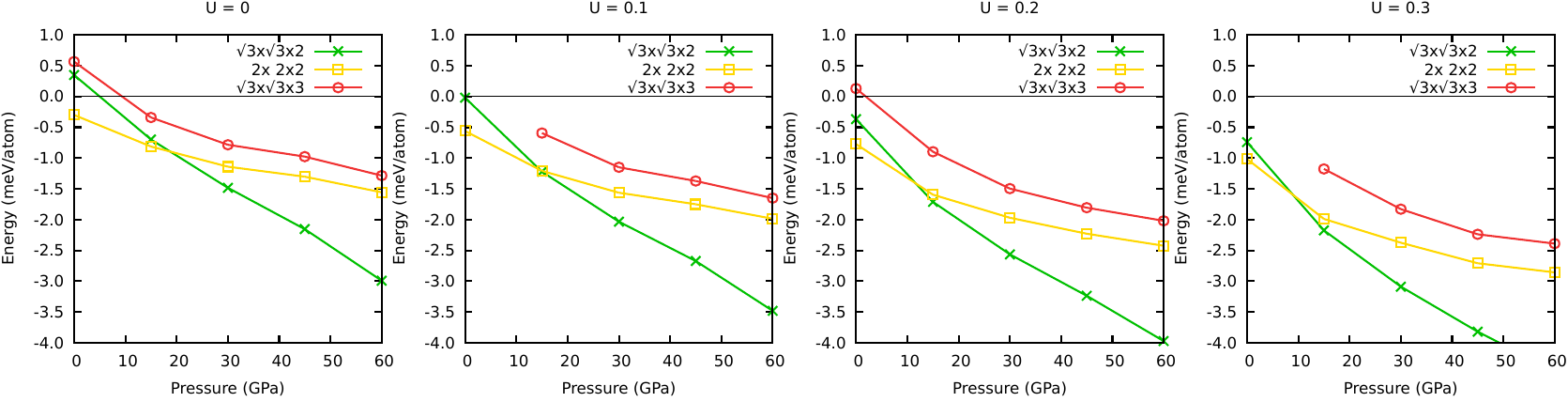}
    \caption{Total energy differences between 3 observed CDW superstructures and the undistorted structure as a function of applied pressure, with Hubbard $U=0,0.1,0.2,0.3$ eV considered in the DFT+U calculation.}
    \label{fig:en-diff-333}
\end{figure}

As pressure goes up to 30 GPa, an additional CDW order is observed with diffraction vector $(1/3,1/3,1/3)$, indicating a $\sqrt{3}\times \sqrt{3}\times 3$ superstructure.
Similarly, utilizing the phonon eigenvectors, the corresponding superstructure is generated with $\vb{q}_1=(1/3,1/3,1/3)$ and its stars, i.e., $\vb{q}_2=(-1/3,-1/3,-1/3)$, $\vb{q}_3=(-1/3,-1/3, 1/3)$, $\vb{q}_4=( 1/3, 1/3,-1/3)$.
The total energy differences between the $\sqrt{3}\times\sqrt{3}\times 3$ superstructure and the undistorted structure with pressure varying from 0 GPa to 60 GPa are calculated, as shown in \cref{fig:en-diff-333}.
However, in calculations without or with Hubbard U considered, the additional $\sqrt{3}\times\sqrt{3}\times3$ structure consistently has higher energy than the previously observed $2\times2\times2$ and $\sqrt{3}\times\sqrt{3}\times2$ structure, which requires further understanding.

\section{Ising model and Monte Carlo simulations}

Again, we can model this disorder using the binary Ising variables +/- where + is an undimerised pair and - is a dimer. We can implement this model by putting the Ising variables (physically represented by two species of dummy atoms within the model) on the midpoints between every pair of trigonal Ge atoms along the c-axis. One third of the dummy atoms represent dimers (-) and two thirds non-dimers (+) in accordance with the average structure. These dummy atoms are ordered by MC using the Hamiltonian in \ref{MC_1} with the following parameter values: \textit{c}$_1$ = 0.3, \textit{c}$_2$ = -0.6, \textit{c}$_3$ = 0.5, \textit{c}$_4$ = 0.4, \textit{c}$_5$ = 0.4 and the MC temperature = 1.0. These values are dimensionless and were chosen based on the match between the diffuse scattering from the resulting atomic model and the experimental scattering. Once the MC has converged, the trigonal Ge atoms are displaced towards the (-) dummy atoms to create dimers. This is done in a second MC step, in which a small displacement in \textit{z} is applied and the energy before and after is calculated using a Lennard-Jones potential, with its minimum is at the $\frac{1}{2}$ of the Ge-Ge dimer distance.

The MC for the $\sqrt{3}\ \times\ \sqrt{3}\ \times$ 2 structure was performed using inhouse code, and the resulting diffuse scattering was calculated using Scatty \cite{Paddison_2018}, whereas the MC and diffuse scattering calculations for the $\sqrt{3}\ \times\ \sqrt{3}\ \times$ 3 structure were performed in the program DISCUS \cite{discuss_1997}.
A schematic of the disorder in both structures is given in Figure 13 and example atomic configurations from the MC simulations (fig. 3(b-c)) are shown  alongside their Fourier transforms (fig. 3(f-g)) in the main text.



\begin{equation}
\begin{split}
    H = \sum_{<i,j>:NN}c_1\sigma_i\sigma_j + \sum_{<i,j>:NNN}c_2\sigma_i\sigma_j + \sum_{<i,j>:4NN}c_3\sigma_i\sigma_j + \\
    \sum_{<i,j>:z-NN}c_4\sigma_i\sigma_j + \sum_{<i,j>:z-NNN}c_5\sigma_i\sigma_j,
    \end{split}
\label{MC_1}
\end{equation}

\begin{figure}
    \centering
    \includegraphics[width=0.35\linewidth]{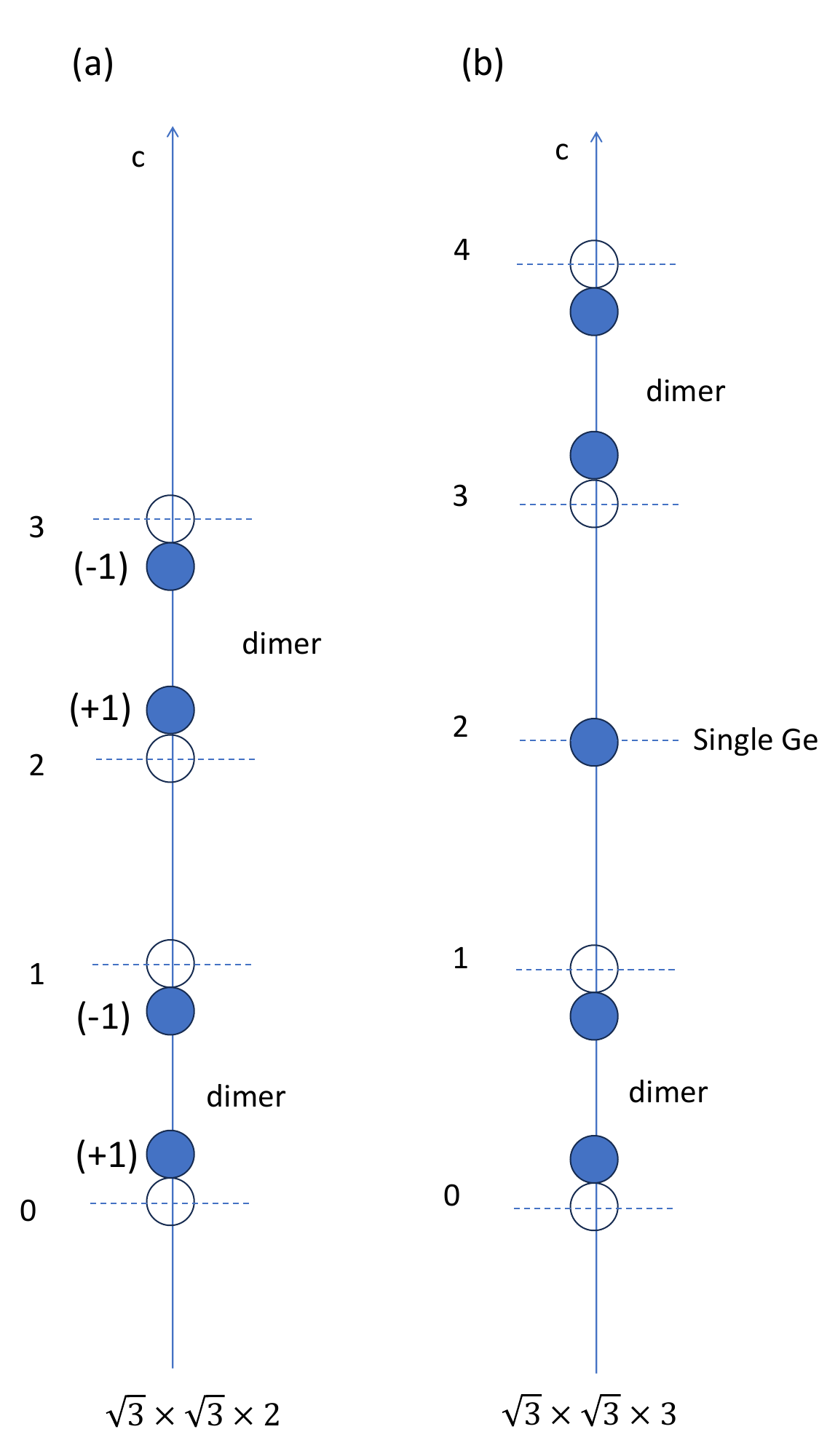}
    \caption{Toy model representing the dimerization of the trigonal Ge$_1$ in the (a) $\sqrt{3}\ \times\ \sqrt{3}\ \times$ 2 and (b) $\sqrt{3}\ \times\ \sqrt{3}\ \times$ 3 orders. +(-)1 stands for upwards (downwards) Ge$_1$ displacement.}
    \label{Fig4_SI}
\end{figure}


\section{\label{app:lg-th}Landau-Ginzburg Theory}
To further investigate the CDW phase transition, we implement the Landau-Ginzburg theory. The relevant order parameters are the phonon fields with momenta $(1/2,0,1/2)$, $(-1/2, 1/2,1/2)$, and $(0,1/2,1/2)$—i.e., the $L$ point for the $2\times 2\times 2$ CDW order, and $(1/3,1/3,1/2)$, $(-1/3, -1/3,-1/2)$—i.e., the $H$ point for the $\sqrt{3}\times \sqrt{3}\times 2$ order, where the momentum is defined according to the nonmagnetic BZ. In the AFM phase, the corresponding momenta become $(1/2,0,0)$, $(-1/2, 1/2,0)$, $(0,1/2,0)$, $(1/3,1/3,0)$, and $(-1/3, -1/3,0)$, corresponding to three $M$ points and two $K$ points. In the following discussion, we use the notation defined according to the AFM BZ.

\subsection{$2\times 2\times 2$ CDW Order}
The $2\times 2\times 2$ CDW order is defined by three nonequivalent $M$ points in the magnetic BZ. To investigate this order, we introduce the following three order parameters:
\begin{aleq}
    \phi_1 \equiv \phi_{M_1}(\vb{q}),\quad \phi_2 \equiv \phi_{M_2}(\vb{q}),\quad \phi_3 \equiv \phi_{M_3}(\vb{q})
\end{aleq}
which correspond to the phonon fields with momenta $M_1 (1/2,0,0)$, $M_2 (0,1/2,0)$, and $M_3 (-1/2, 1/2,0)$, respectively. These fields form the $M1-$ irreducible representation (irrep) of the little group $D_{2h}$ at $M$.

Under a symmetry operation $g$ in SG 192, the phonon field operators transform according to the following relations:
\begin{aleq}
    g \phi_i g^{-1} \eq \sum_j \phi_j \bqty{D(g)}_{ij}.
\end{aleq}
As discussed in \cref{app:cryst}, due to the A-type AFM configuration, the unitary part of the space group of FeGe is now SG 192 $P6/mcc$. Specifically, for the generators $g$ of SG 192 (in addition to time-reversal symmetry $\mathcal{T}$ and translational symmetry $T_{\vb{R}}$), the matrix representations $D(g)$ under the basis of phonon field operators from the $M1-$ irrep are:
\begin{aleq}
    D(C_{3,001}) \eq \mqty( 0&1&0\\0&0&1\\1&0&0), \quad
    &D(C_{2,001}) \eq \mqty(1&0&0\\0&1&0\\0&0&1), \quad
    &D(\tilde{C}_{2,110}) \eq \mqty(0&1&0\\0&0&1\\1&0&0), \\
    D(\mathcal{P}) \eq \mqty(-1&0&0\\0&-1&0\\0&0&-1), \quad
    &D(\mathcal{T}) \eq \mqty(1&0&0\\0&1&0\\0&0&1), \quad
    &D(T_{\vb{R}}) \eq \mqty(e^{iM_1\cdot\vb{R}}&0&0\\ 0&e^{iM_2\cdot\vb{R}}&0\\0&0&e^{iM_3\cdot\vb{R}}),
\end{aleq}
where $\tilde{C}_{2,110}\equiv\Bqty{C_{2,110}|\pqty{0,0,1/2}}$.

Additionally, we have the following relation:
\begin{aleq}
    \phi_i^\dag \eq \phi_i,\quad i = 1, 2, 3.
\end{aleq}

The free energy $\mathcal{L}_{\phi}$ of the $\phi$ fields can be expanded to the quartic term as:
\begin{aleq}
    \mathcal{L}_\phi \eq m_\phi \sum_i \phi_i^2 + u_\phi \pqty{\sum_i \phi_i^2}^2,
\end{aleq}
where the cubic term $\prod_i \phi_i$ is odd under inversion $\mathcal{P}$ and therefore forbidden. This suggests that the $2\times 2\times 2$ CDW phase transition is likely second order. However, a first-order transition could still occur if higher-order terms, such as a sixth-order term, are introduced.

\subsection{$\sqrt{3}\times \sqrt{3}\times 2$ CDW Order}
Similarly, the $\sqrt{3}\times \sqrt{3}\times 2$ CDW order is defined by two nonequivalent $K$ points in the magnetic BZ. To investigate this order, we introduce the following two order parameters:
\begin{aleq}
    \psi_1 \equiv \psi_{K_1}(\vb{q}),\quad \psi_2 \equiv \psi_{K_2}(\vb{q}),
\end{aleq}
which correspond to the phonon fields with momenta $K_1 (1/3,1/3,0)$ and $K_2 (-1/3,-1/3,0)$, respectively. These form the $K2$ irreducible representation (irrep) of the little group $D_{3h}$ at $K$.

For the generators $g$ of SG 192 (along with time-reversal symmetry $\mathcal{T}$ and translational symmetry $T_{\vb{R}}$), the matrix representations $D(g)$ under the basis of phonon field operators from the $K2$ irrep are given by:
\begin{aleq}
    D(C_{3,001}) \eq \mqty(1&0\\0&1), \quad
    &D(C_{2,001}) \eq \mqty(0&-1\\-1&0), \quad
    &D(\tilde{C}_{2,110}) \eq \mqty(1&0\\0&1), \\
    D(\mathcal{P}) \eq \mqty(0&1\\1&0), \quad
    &D(\mathcal{T}) \eq \mqty(0&1\\1&0), \quad
    &D(T_{\vb{R}}) \eq \mqty(e^{iK_1\cdot \vb{R}} & 0 \\ 0 & e^{iK_2\cdot \vb{R}}).
\end{aleq}

Additionally, we have the following relation:
\begin{aleq}
    \psi_2^\dag \eq \psi_1.
\end{aleq}
Therefore, it is sufficient to consider only $\psi = \psi_1$.

The free energy $\mathcal{L}_{\psi}$ of the $\psi$ fields can be expanded to the quartic term as:
\begin{aleq}
    \mathcal{L}_\psi \eq m_\psi \abs{\psi}^2 + u_\psi \abs{\psi}^4,
\end{aleq}
where the cubic term is odd under the mirror-$z$ operation $m_z\equiv\mathcal{P} C_{2z}$ and therefore forbidden, indicating that the $\sqrt{3}\times \sqrt{3}\times 2$ CDW phase transition is likely to be second order. However, a first-order transition could still occur if higher-order terms, such as a sixth-order term, are introduced.

\subsection{$\sqrt{3}\times \sqrt{3} \times 3$ CDW order}
At even higher pressure (over 30 GPa in experiment), an additional CDW order is observed with XRD peaks at $(1/3,1/3,1/3)$, indicating a cell tripling along $\hat{\vb{c}}$ direction.
The $\sqrt{3}\times \sqrt{3}\times 3$ CDW order is defined by four nonequivalent $\bar{K}$ points in the non-magnetic BZ. To investigate this order, we introduce the following four order parameters:
\begin{aleq}
    \varphi_1 \equiv \varphi_{\bar{K}_1}(\vb{q}), \quad 
    \varphi_2 \equiv \varphi_{\bar{K}_2}(\vb{q}), \quad 
    \varphi_3 \equiv \varphi_{\bar{K}_3}(\vb{q}), \quad 
    \varphi_4 \equiv \varphi_{\bar{K}_4}(\vb{q}),
\end{aleq}
which correspond to the phonon fields with momenta 
\begin{aleq}
    \bar{K}_{1} \eq ( 1/3,  1/3,  1/3), \\
    \bar{K}_{2} \eq (-1/3, -1/3,  1/3), \\
    \bar{K}_{3} \eq ( 1/3,  1/3, -1/3), \\
    \bar{K}_{4} \eq (-1/3, -1/3, -1/3).
\end{aleq}
These form the $P1$ irreducible representation (irrep) of the little group $C_{3v}$ at $\bar{K}$.

Since the space group of the new $\sqrt{3}\times\sqrt{3}\times 3$ order remains unknown, we assume it with highest symmetry of possible subgroup induced by irrep $P1$, $P6/mmm$.
For the generators $g$ of SG 191 (along with time-reversal symmetry $\mathcal{T}$ and translational symmetry $T_{\vb{R}}$), the matrix representations $D(g)$ under the basis of phonon field operators from the $P1$ irrep are given by:
\begin{aleq}
    D(C_{3,001}) \eq \sigma_0 \otimes \sigma_0, \quad
    & D(M_{1\bar{1}0}) \eq \sigma_0 \otimes \sigma_x, \quad
    & D(C_{2,001}) \eq \sigma_0 \otimes \sigma_x, \\
    D(\mathcal{P}) \eq \sigma_x \otimes \sigma_x, \quad
    & D(\mathcal{T}) \eq \sigma_x \otimes \sigma_x, \quad
    & D(T_{\vb{R}}) \eq \mathrm{diag}( e^{i \bar{K}_1 \cdot \vb{R}}, e^{i \bar{K}_2 \cdot \vb{R}}, e^{i \bar{K}_3 \cdot \vb{R}}, e^{i \bar{K}_4 \cdot \vb{R}})
\end{aleq}
Additionally, we have the following relation:
\begin{aleq}
    \varphi_1^{\dag} \eq \varphi_4; \quad 
    &\varphi_2^{\dag} \eq \varphi_3
\end{aleq}
The free energy $\mathcal{L}_{\varphi}$ of the $\varphi$ fields can be expanded to the quartic term as:
\begin{aleq}\label{eq:lg_333}
    \mathcal{L}_{\varphi} \eq m_{\varphi} \sum_i \abs{\varphi_i}^2 + u_{\varphi,1} \sum_i \pqty{\varphi_i^3 + \varphi_i^{\dag,3}} + u_{\varphi,2} \sum_i \abs{\varphi_i}^4
\end{aleq}
The symmetry allowed cubic terms in \cref{eq:lg_333} shows that the $\bar{K}$ phonon condensation can lead to a first-order phase transition, indicating that the phase transition observed with pressure between $\sqrt{3}\times\sqrt{3}\times2$ and $\sqrt{3}\times\sqrt{3}\times3$ could be of first order, similar to the CDW in ScV$_6$Sn$_6$ \cite{Hu2023KagomeA}.

\subsection{Phase Transition Between Two CDW Phases}

Experimentally, a phase transition between the $2\times 2 \times 2$ CDW order and the $\sqrt{3}\times \sqrt{3} \times 2$ CDW order has been observed as a function of applied pressure. To theoretically describe this phase transition, we consider the coupling between the order parameters associated with the two distinct CDW phases. The symmetry-allowed free energy for this coupling can be expressed as:
\begin{aleq}
    \mathcal{L}_{\phi \psi} \eq u_{\phi\psi} \pqty{\sum_i \phi_i^2} \abs{\psi}^2.
\end{aleq}

\begin{figure}[b]
    \centering
    \includegraphics[width=0.8\linewidth]{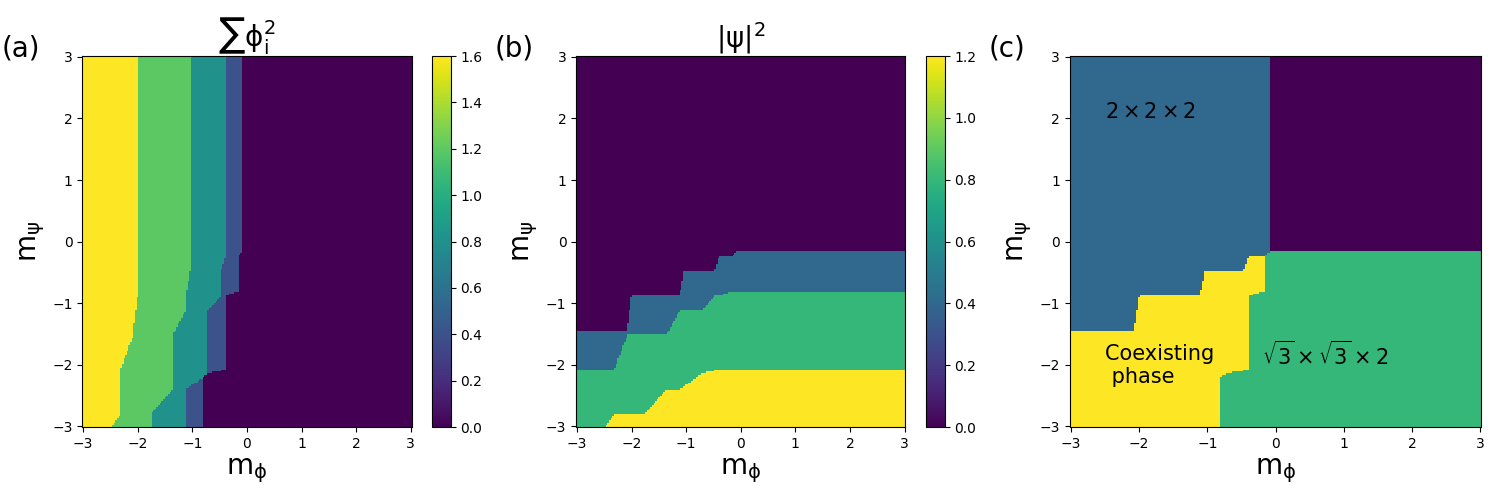}
    \caption{Amplitudes of the order parameters for the $2\times 2 \times 2$ CDW (a) and $\sqrt{3}\times \sqrt{3}\times 2$ CDW (b) phases, respectively. (c) Phase diagram derived from the Landau-Ginzburg free energy.}
    \label{fig:lg_phase_diagram}
\end{figure}

The total free energy can thus be written as:
\begin{aleq}
 \mathcal{L} =  \mathcal{L}_{\phi} + \mathcal{L}_{\psi} +  \mathcal{L}_{\phi \psi},
 \label{eq:lg_free_energy}
\end{aleq}
where, without loss of generality, we assume $u_\phi = u_\psi = 1$ and $u_{\phi\psi} = 0.5$. By solving the Landau-Ginzburg equations derived from \cref{eq:lg_free_energy} for varying the values of $m_\phi$ and $m_\psi$, we obtain the phase diagram depicted in \cref{fig:lg_phase_diagram}. This phase diagram illustrates that, as $m_\phi$ and $m_\psi$ are tuned, a phase transition occurs from a $2 \times 2 \times 2$ CDW phase to a phase where both CDW orders coexist, and eventually to a $\sqrt{3} \times \sqrt{3} \times 2$ CDW phase.

\clearpage

\clearpage
\bibliographystyle{apsrev4-1}
\bibliography{ref}